\documentclass[aps,superscriptaddress,eqsecnum,nofootinbib,showpacs,preprintnumbers,twocolumn]{revtex4-1}
\usepackage{longtable}
\usepackage{bm}
\usepackage{relsize}
\usepackage{amsfonts}
\usepackage{amsmath}
\usepackage{amssymb,epsf}
\usepackage{latexsym}
\usepackage{graphicx,epsfig}
\usepackage{amssymb}
\usepackage{float}
\usepackage{subfigure}
\usepackage{epstopdf}
\usepackage{dcolumn}
\usepackage{psfrag}
\usepackage{wrapfig}
\usepackage{makeidx}
\usepackage{epsf}
\usepackage{color}
\usepackage{multirow}
\usepackage{mathtools}

\def\be{\begin{equation}}
\def\ee{\end{equation}}
\def\bea{\begin{eqnarray}}
\def\eea{\end{eqnarray}}

\begin{document}

\title{Gravitationally induced particle production in scalar-tensor $f(R,T)$ gravity}

\author{Miguel A. S. Pinto}
\email{mapinto@fc.ul.pt}
\affiliation{Instituto de Astrofísica e Ciências do Espaço, Faculdade de Ciências da Universidade de Lisboa, Edifício C8, Campo Grande, P-1749-016 Lisbon, Portugal}
\affiliation{Departamento de F\'{i}sica, Faculdade de Ci\^{e}ncias da Universidade de Lisboa, Edifício C8, Campo Grande, P-1749-016 Lisbon, Portugal}

\author{Tiberiu Harko}
\email{tiberiu.harko@aira.astro.ro}
\affiliation{Department of Theoretical Physics, National Institute of Physics
and Nuclear Engineering (IFIN-HH), Bucharest, 077125 Romania,}
\affiliation{Department of Physics, Babes-Bolyai University, Kogalniceanu Street,
	Cluj-Napoca 400084, Romania,}
\affiliation{Astronomical Observatory, 19 Ciresilor Street,
	Cluj-Napoca 400487, Romania,}

\author{Francisco S. N. Lobo}
\email{fslobo@fc.ul.pt}
\affiliation{Instituto de Astrofísica e Ciências do Espaço, Faculdade de Ciências da Universidade de Lisboa, Edifício C8, Campo Grande, P-1749-016 Lisbon, Portugal}
\affiliation{Departamento de F\'{i}sica, Faculdade de Ci\^{e}ncias da Universidade de Lisboa, Edifício C8, Campo Grande, P-1749-016 Lisbon, Portugal}

\date{\today}

\begin{abstract}
We explore the possibility of gravitationally generated particle production in the scalar-tensor representation of $f(R,T)$ gravity. Due to the explicit nonminimal curvature-matter coupling in the theory, the divergence of the matter energy-momentum tensor does not vanish. We explore the
physical and cosmological implications of this property by using the formalism of irreversible thermodynamics of open systems in the presence of matter creation/annihilation. The particle creation rates, pressure, temperature evolution and the expression of the comoving entropy are obtained in a covariant formulation and discussed in detail. Applied together with the gravitational field equations, the thermodynamics of open systems lead to a generalization of the standard $\Lambda$CDM cosmological paradigm, in which the particle creation rates and pressures are effectively considered as components of the cosmological fluid energy-momentum tensor. We also consider specific models, and compare the scalar-tensor $f(R,T)$ cosmology with the $\Lambda$CDM scenario and the observational data for the Hubble function. The properties of the particle creation rates, of the creation pressures, and entropy generation through gravitational matter production are further investigated in both the low and high redshift limits.
\end{abstract}

\pacs{04.50.Kd, 04.20.Cv}

\maketitle


\section{Introduction}\label{sec:intro}

Einstein's General Theory of Relativity (GR) is undoubtedly one of the most extraordinary theories ever conceived by the human mind \cite{Ein}. Its mathematical representation is given by the Einstein gravitational field equations,  $G_{\mu \nu} = R_{\mu \nu} - (1/2)Rg_{\mu \nu} = \kappa ^2 T_{\mu \nu}$, where $R_{\mu \nu}$ is the contraction of the Riemann curvature tensor, $R_{\mu \nu}=R^{\kappa}_{\mu \kappa \nu}$, $R$ is the Ricci scalar $R=R_{\mu}^{\mu}$, $T_{\mu \nu}$ is the matter energy-momentum tensor, and $\kappa ^2
=8\pi G/c^4$ is the gravitational coupling constant. The gravitational field equations can also be derived from a variational principle, introduced by Hilbert \cite{Hilb}, thus making them fully consistent gravitationally. GR had an amazing success in explaining gravitational dynamics at the level of the Solar System, including the precession of the perihelion of the planet Mercury, the deflection of light, gravitational redshift, or the Shapiro delay effect \cite{Will}.  One of its outstanding predictions, namely, the existence of gravitational waves (GWs) has been directly confirmed by the LIGO-VIRGO collaboration \cite{LIGOScientific:2016aoc}. This amazing discovery has opened a new window to test the nature of gravity and paved the way for a new era in astronomy, astrophysics, and fundamental physics. In fact, ESA selected the science theme -- The Gravitational Universe -- and the spaceborne observatory of GWs, LISA, as the goal for the third large mission (L3) in its Cosmic Vision Program.

However, GR is currently facing many theoretical and experimental challenges.  High precision data obtained from the observations of the Type Ia supernovae  has confirmed with startling evidence that the Universe is presently in a phase of a de Sitter type accelerated expansion \cite{Ri98,Ri98-1,Ri98-2,Ri98-3, Ri98-4,Hi,Ri98-5}. These important observations have led to a plethora of observational and theoretical works, attempting to explain, and understand, the present, as well as the past,  cosmological dynamics (for a recent review of the cosmic acceleration problem see \cite{Ri98-6}). On the other hand, the Planck satellite observations of the Cosmic Microwave
Background \cite{Pl}, in conjunction with the studies of the Baryon Acoustic Oscillations \cite{Da1,Da2,Da3}, have also confirmed the late-time cosmic acceleration. However, to elucidate these important discoveries an essential modification in our present day understanding of the gravitational interaction is needed. Indeed, to solve some of the outstanding theoretical and observational problems of modern cosmology, one postulates the existence of an exotic cosmic fluid, which possesses a repulsive character at large scales, denoted dark energy \cite{Copeland:2006wr, PeRa03, Pa03,DE1,DE2,DE3,DE4,quint1b,quint2b,quint3b,quint4b, quint5b}.

Another fundamental problem in astrophysics and cosmology is represented by the dark matter problem (see \cite{DMR1,DMR2,DMR3} for reviews on the existence of dark matter, its properties, and on the recent for its search). On galactic and extragalactic scales the presence of dark matter is necessary for a (possible) explanation of two basic astrophysical/astronomical observations, namely, the dynamics of the galactic rotation curves, and the virial mass problem in clusters of galaxies, respectively. The detailed astronomical observations of the galactic rotation curves \cite{Sal, Bin, Per, Bor} indicate that Newtonian gravity, and also GR, cannot explain galactic dynamics. The properties of the galactic rotation curves and the missing mass problem in clusters of galaxies are usually explained by postulating the existence of another dark (invisible) form of matter, and whose interaction with baryonic matter is only gravitational.  Dark matter is assumed to reside around galaxies, where it forms a spherically symmetric halo. Dark matter is usually described as a pressureless and cold cosmic fluid (for reviews of the candidates for dark matter particle see \cite{Ov, Ov1,Ov2,Ov3}).

Therefore, in our present day understanding of the Universe, the local dynamics, and the global expansion are controlled by two major components, cold dark matter, and dark energy, respectively. Hence, baryonic matter plays a negligible role in the late time cosmic expansion. Probably the simplest theoretical model, which can fully explain the late de Sitter type expansion, is obtained from the Einstein field equations that also include the cosmological constant $\Lambda$, introduced by Einstein in 1917 \cite{Einb} in order to obtain a static curved model of the Universe. 
The extension of the Einstein field equations through the addition of $\Lambda$ represents the basic theoretical and mathematical tool of the standard present day cosmological paradigm, the $\Lambda$CDM ($\Lambda$ Cold Dark Matter) model. Even though the $\Lambda$CDM model fits very well the observational data \cite{C1,C2,C3, C4}, it possesses theoretical problems (for reviews and detailed discussions on the cosmological constant problem see, for example,  \cite{W1b,W2b,W3b}).

However, recently the $\Lambda $CDM model is also faced with some other important observational problems. The ``Hubble tension'' is perhaps the most important of these problems. It originated from the differences obtained for the numerical values of the Hubble constant, $H_0$ by using different observational methods. Thus, the determinations of $H_0$ by the Planck satellite by using measurements of the Cosmic Microwave Background  \cite{C4}, do not match with the values measured by using observations in the local Universe \cite{M1,M2,M3}. For example, the SH0ES determinations of $H_0$ give the value $H_0 = 74.03 \pm 1.42$ km/s/Mpc \cite{M1}. On the other hand, the analysis of the CMB by using the Planck satellite data gives $H_0 = 67.4 \pm 0.5$ km/s/Mpc \cite{C3}, with this value differing from the SH0ES result by $\sim  5\sigma$.

Therefore, the pursuit for alternative descriptions of the cosmological expansion, and of the nature of dark matter represents a fundamental undertaking for present day astrophysics and cosmology. One of the interesting possibilities for solving the cosmological mysteries is to move beyond the framework of standard GR, by resorting to modified gravity theories.  Indeed, the difficulty in explaining specific observations, the incompatibility with other well-established theories and the lack of uniqueness, might be indicative of a need for new gravitational physics \cite{Harko:2018ayt}. Hence, the assumption that the gravitational force itself changes on cosmological scales  could represent a very promising direction of investigation of cosmological phenomena \cite{DeFelice:2010aj,Clifton:2011jh,Sotiriou:2008rp, Od0, Avelino:2016lpj, Od}.

An interesting class of modified gravity theories are represented by theories that involve curvature-matter couplings
\cite{Harko:2010mv,Harko:2012hm,Haghani:2013oma,Harko:2014gwa,Harko:2014sja,Harko:2014aja,Harko:2018gxr,Harko:2020ibn}, in which matter plays a more important role than in standard GR, through its direct effect on geometry. In this work, we consider the cosmological implications of a specific theory of modified gravity, with an explicit curvature-matter coupling, namely, the $f(R,T)$ gravity theory \cite{Harko:2011kv}.  An important feature of this theory, as well as of all theories with geometry-matter coupling is that the covariant divergence of the energy-momentum tensor does not vanish \cite{Koivisto:2005yk,Bertolami:2007gv}, and hence $T_{\mu\nu}$ is no longer conserved. Constraints on the viability of these theories have also been considered in the literature \cite{Alvarenga:2013syu}. In fact, the non-conservation leads to the possibility of an energy/matter transfer from the gravitational field to ordinary matter, and therefore it may involve matter creation processes.

Hence, inspired by the basic principles of the thermodynamic of open systems, we investigate the possibility of gravitationally induced particle production in the scalar-tensor representation of $f(R,T)$ gravity \cite{daSilva:2021dsq,Goncalves:2021vci}. We explore the physical and cosmological implications of the non-conservation of the theory, which we associate with particle creation, by using the formalism of irreversible thermodynamics of open system, in the presence of matter creation \cite{Pri0,Pri}. A covariant formulation of the irreversible thermodynamics was developed in \cite{Calvao:1991wg}. In fact, irreversible thermodynamics and thermodynamics of open systems is a widely studied field, since it is useful in various applications \cite{Harko:2014pqa,Harko:2015pma,Su,Iv1,Iv2,Ba,Si,HaH,GoSa,Hama, Harko:2021bdi}.
We will obtain the particle creation rate and creation pressure in the scalar-tensor representation of $f(R,T)$ gravity, and we will discuss in detail their properties, and implications. We also consider specific cosmological models, and compare the predictions of the theory with the $\Lambda$CDM scenario.

This article is organised in the following manner: In Section~\ref{sec:fRTintro}, we outline the geometrical and the scalar-tensor representations of $f(R,T)$ gravity, and present the modified field equation. In Section~\ref{sec:cosmo}, the generalized Friedmann equations are presented and we explore the physical and thermodynamical implications in the framework of the thermodynamics of open systems, by assuming that the nonconservation of the energy-momentum describes an irreversible matter creation process.
In Section~\ref{sec:models}, we consider several cosmological models by specifying the functional form of the potential $V(\varphi,\psi)$, and we perform a detailed comparison of the theoretical models  with the cosmological observations, and with the predictions of the standard $\Lambda$CDM scenario.
We summarize and discuss our results in Section~\ref{sec:conclusion}.

\section{Field equations of $f\left(R,T\right)$ gravity}\label{sec:fRTintro}

In the present section, we introduce the action principle for the $f(R,T)$ gravity theory, and we present its scalar-tensor representation with the help of two independent scalar fields.

\subsection{Geometrical Representation}

We assume that the action for $f\left(R,T\right)$ gravity is given by the following expression \cite{Harko:2011kv}
\begin{equation}\label{eq:fRTaction-original}
    S = \frac{1}{2\kappa^2} \int_{\Omega}\sqrt{-g} \, f(R,T) d^4 x+ \int_{\Omega} \sqrt{-g} \, \mathcal{L}_m d^4 x,
\end{equation}
where $\kappa^2=8\pi G$, $G$ is the universal gravitational constant, $\Omega$ is the 4-dimensional spacetime manifold on which a set of coordinates $x^\mu$ is defined, $g$ is the determinant of the metric tensor $g_{\mu\nu}$. $f\left(R,T\right)$ is an arbitrary well-behaved function of the Ricci scalar $R=g^{\mu\nu}R_{\mu\nu}$, where $R_{\mu\nu}$ is the Ricci tensor, and the trace of the energy-momentum tensor $T=g^{\mu\nu}T_{\mu\nu}$ $T_{\mu\nu}$, which is defined in terms of the variation of the matter Lagrangian $\mathcal L_m$ as
\begin{equation}
T_{\mu\nu}=-\frac{2}{\sqrt{-g}}\frac{\delta\left(\sqrt{-g}\mathcal L_m\right)}{\delta g^{\mu\nu}}.
\end{equation}
Throughout this work we adopt a system of geometrized units, where $G=c=1$, and therefore $\kappa^2=8\pi$.

By varying the action \eqref{eq:fRTaction-original} with respect to the metric tensor $g_{\mu\nu}$, we obtain the following gravitational field equations of $f(R,T)$ gravity (see Ref.~\cite{Harko:2011kv} for details)
\begin{equation}\label{eq:fields-original}
\begin{multlined}
    f_R R_{\mu\nu}-\frac{1}{2}g_{\mu\nu}f(R,T) + \left(g_{\mu\nu}\square-\nabla_\mu\nabla_\nu\right)f_R \\
    = \kappa^2 T_{\mu\nu}-f_T (T_{\mu\nu}+\Theta_{\mu\nu}),
\end{multlined}
\end{equation}
where $f_R$ and $f_T$ denote partial derivatives of $f$ with respect to $R$ and $T$, respectively, $\nabla_\mu$ is the covariant derivative and $\square\equiv\nabla^\sigma\nabla_\sigma$ is the D’Alembert operator, both defined in terms of the metric tensor $g_{\mu\nu}$. The tensor $\Theta_{\mu\nu}$ is defined as
\begin{equation}\label{eq:Theta-varT}
    \Theta_{\mu\nu}\equiv g^{\rho\sigma}\frac{\delta T_{\rho\sigma}}{\delta g^{\mu\nu}}.
\end{equation}
By taking the divergence of Eq.~\eqref{eq:fields-original} and using the geometric identity $\left(\square\nabla_\nu-\nabla_\nu\square\right)f_R=R_{\mu\nu}\nabla^\mu f_R$, one finds that the conservation equation for $f\left(R,T\right)$ gravity is the following
\begin{equation}\label{eq:conserv-general}
\begin{multlined}
    (\kappa^2-f_T)\nabla^\mu T_{\mu\nu}=\left(T_{\mu\nu}+\Theta_{\mu\nu}\right)\nabla^\mu f_T \\
    +f_T\nabla^\mu\Theta_{\mu\nu}+f_R \nabla^\mu R_{\mu\nu}-\frac{1}{2}g_{\mu\nu}\nabla^\mu f.
\end{multlined}
\end{equation}
As we can see the covariant divergence of the energy-momentum tensor of matter does not vanish. We interpret this result as an exchange of energy and momentum between geometry and matter, as a consequence of the geometry-matter coupling encapsulated in the trace $T$ of the energy-momentum tensor of matter. Next we consider the scalar-tensor representation of this theory, which will be used throughout this work.

\subsection{Scalar-Tensor Representation}\label{subsec:scalar-tensor}

$f\left(R,T\right)$ gravity can be rewritten in a dynamically equivalent scalar-tensor representation with two scalar fields \cite{Goncalves:2021vci}. First, one introduces two auxiliary fields $\alpha$ and $\beta$ and rewrites the action \eqref{eq:fRTaction-original} in the form
\begin{eqnarray}\label{eq:STaction-intro}
    S &=& \frac{1}{2\kappa^2} \int_{\Omega}\sqrt{-g} \big[f(\alpha,\beta)+ (R-\alpha)f_\alpha
		\nonumber \\
    && +(T-\beta)f_\beta\big] d^4 x  + \int_{\Omega} \sqrt{-g}\mathcal{L}_m  d^4 x ,
\end{eqnarray}
where the subscripts $\alpha$ and $\beta$ in $f$ denote its partial derivatives with respect to these variables, respectively.

We define the two scalar fields $\varphi$ and $\psi$ and a scalar interaction potential $V\left(\varphi,\psi\right)$ in the forms
 \begin{equation}\label{eq:varphi&psi}
     \varphi\equiv\frac{\partial f}{\partial R} ,\qquad
    \psi\equiv\frac{\partial f}{\partial T},
 \end{equation}
\begin{equation}\label{eq:potential}
    V(\varphi,\psi) \equiv -f(\alpha,\beta)+ \varphi \alpha + \psi \beta ,
\end{equation}
so that the action \eqref{eq:STaction-intro} is rewritten in the equivalent scalar-tensor representation as
\begin{equation}\label{eq:STaction}
    \begin{split}
    S = \frac{1}{2\kappa^2} \int_{\Omega} \sqrt{-g} \left[\varphi R+\psi T - V(\varphi, \psi)\right]d^4 x \\
    + \int_{\Omega} \sqrt{-g} \mathcal{L}_m d^4 x .
    \end{split}
\end{equation}

The action \eqref{eq:STaction} depends on three independent variables, the metric $g_{\mu\nu}$ and the two scalar fields $\varphi$ and $\psi$. Varying the action with respect to the metric $g_{\mu\nu}$ yields the following gravitational field equations
\begin{equation}\label{eq:fields}
    \begin{multlined}
      \varphi R_{\mu\nu}-\frac{1}{2}g_{\mu\nu}\left(\varphi R + \psi T - V\right)\\-(\nabla_\mu\nabla_\nu-g_{\mu\nu}\square)\varphi = \kappa^2 T_{\mu\nu} -\psi (T_{\mu\nu} + \Theta_{\mu\nu}).
      \end{multlined}
\end{equation}
This field equation can also be obtained directly from Eq.~\eqref{eq:fields-original} by using the definitions shown in Eqs.~\eqref{eq:varphi&psi} and \eqref{eq:potential}, with $\alpha=R$ and $\beta=T$. Taking the variation of Eq.~\eqref{eq:STaction} with respect to the scalar fields $\varphi$ and $\psi$ yield the following relations
\begin{equation}\label{eq:Vphi}
    V_{\varphi} = R,
\end{equation}
\begin{equation}\label{eq:Vpsi}
    V_{\psi} = T,
\end{equation}
respectively, where the subscripts in $V_\varphi$ and $V_\psi$ denote the derivatives of the potential $V(\varphi,\psi)$ with respect to the variables $\varphi$ and $\psi$, respectively.\par

Additionally, using the definitions \eqref{eq:varphi&psi} and \eqref{eq:potential}, and the geometrical result $\nabla^\mu\left[R_{\mu\nu}-\left(1/2\right)Rg_{\mu\nu}\right]=0$, the conservation equation for $f\left(R,T\right)$ gravity in the scalar-tensor representation becomes
\begin{equation}\label{eq:conserv-general2}
\begin{multlined}
    (\kappa^2-\psi)\nabla^\mu T_{\mu\nu}=\left(T_{\mu\nu}+\Theta_{\mu\nu}\right)\nabla^\mu\psi
    \\
    +\psi\nabla^\mu\Theta_{\mu\nu}-\frac{1}{2}g_{\mu\nu}\left[R\nabla^\mu\varphi+\nabla^\mu\left(\psi T-V\right)\right].
\end{multlined}
\end{equation}

The results obtained in this subsection will be used to explore cosmological solutions in the next section.

\section{Cosmological evolution}\label{sec:cosmo}

In order to consider the cosmological evolution in the scalar-tensor formulation of $f(R,T)$ gravity, we obtain first the generalized Friedmann equations corresponding to the Friedmann-Lemaitre-Robertson-Walker metric. Then, after a brief presentation of the thermodynamic of open systems, we proceed to apply it systematically to the cosmological models in the $f(R,T)$ gravity theory. The particle creation rates, the creation pressures, as well as the entropy and temperature evolutions are considered in detail.
\subsection{Spacetime geometry, and generalized Friedmann equations}

In this work, we assume that the Universe is described by an homogeneous and isotropic flat FLRW spacetime metric, given in spherical coordinates $(t,r,\theta,\phi)$ by
\begin{equation}\label{eq:FLRW-metric}
    ds^2 = -dt^2+a^2(t)\left[ dr^2+r^2\left(d\theta^2+\sin^2\theta d\phi^2\right)\right],
\end{equation}
where $a(t)$ is the scale factor. In addition to this, we also assume that matter is described by a perfect fluid:
\begin{equation}\label{eq:em-fluid}
    T_{\mu\nu}=(\rho+p)u_\mu u_\nu +p g_{\mu\nu},
\end{equation}
where $\rho$ is the energy density, $p$ is the isotropic pressure, and $u^\mu$ is the fluid 4-velocity satisfying the normalization condition $u_\mu u^\mu=-1$. Taking the matter Lagrangian to be $\mathcal L_m=p$ \cite{Bertolami:2008ab}, the tensor $\Theta_{\mu\nu}$ takes the form
\begin{equation}\label{eq:Theta-fluid}
    \Theta_{\mu\nu}=-2T_{\mu\nu}+p g_{\mu\nu}.
\end{equation}

To preserve the homogeneity and isotropy of the solution, all physical quantities are assumed to depend solely on the time coordinate $t$, i.e., $\rho=\rho\left(t\right)$, $p=p\left(t\right)$, $\varphi=\varphi\left(t\right)$, and $\psi=\psi\left(t\right)$. Under these assumptions, one obtains two independent field equations from Eq.~\eqref{eq:fields}, namely, the modified Friedmann equation and the modified Raychaudhuri equation, which take the following forms
\begin{equation}\label{eq:tt}
    \dot{\varphi}\left(\frac{\dot{a}}{a}\right) + \varphi \left( \frac{\dot{a}}{a} \right)^2  = \frac{8 \pi}{3}\rho + \frac{\psi}{2}\left( \rho - \frac{1}{3}p\right)+ \frac{1}{6} V,
\end{equation}
\begin{equation}\label{eq:rr}
\begin{split}
   \ddot{\varphi}+ 2\dot{\varphi}\left(\frac{\dot{a}}{a}\right) +\varphi\left( \frac{2\ddot{a}}{a} + \frac{\dot{a}^{2}}{a^{2}} \right)  = -8\pi p  \\+ \frac{\psi}{2}\left(\rho-3p\right) + \frac{1}{2} V,
\end{split}
\end{equation}
respectively, where overdots denote derivatives with respect to time. Furthermore, the equations of motion for the scalar fields $\varphi$ and $\psi$ from Eqs.~\eqref{eq:Vphi} and \eqref{eq:Vpsi} become
\begin{equation}\label{eq:Vphicosmo}
    V_{\varphi} =R= 6\left( \frac{\ddot{a}}{a}+ \frac{\dot{a}^{2}}{a^{2}}\right),
\end{equation}
\begin{equation}\label{eq:Vpsicosmo}
    V_{\psi} =T= 3p-\rho,
\end{equation}
respectively. Finally, the conservation equation from Eq.~\eqref{eq:conserv-general2} in this framework takes the form
\begin{equation}\label{eq:conserv-total}
\begin{multlined}
 \dot{\rho}  +   3(\rho+p)\left(\frac{\dot{a}}{a}\right)= \frac{3}{8\pi} \Bigg\{ \dot{\varphi}\left(\frac{\ddot{a}}{a}+\frac{\dot{a}^2}{a^2}-\frac{1}{6}V_\varphi\right) \\
    -\dot{\psi}\left(\frac{1}{2}\rho - \frac{1}{6}p+\frac{1}{6}V_\psi\right) -\psi\left[\frac{\dot{a}}{a}(\rho+p)+\frac{1}{2}\dot{\rho} - \frac{1}{6}\dot{p}\right]  \Bigg\}.
\end{multlined}
\end{equation}

The system of Eqs.~\eqref{eq:tt}--\eqref{eq:conserv-total} forms a system of five equations from which only four are linearly independent. To prove this feature, one can take the time derivative of Eq.~\eqref{eq:tt}, use Eqs.~\eqref{eq:Vphicosmo} and \eqref{eq:Vpsicosmo} to eliminate the partial derivatives $V_\varphi$ and $V_\psi$, use the conservation equation in Eq.~\eqref{eq:conserv-total} to eliminate the time derivative $\dot\rho$, and use the Raychaudhuri equation in Eq.~\eqref{eq:rr} to eliminate the second time derivative $\ddot a$, thus recovering the original equation. Thus, one of these equations can be discarded from the system without loss of generality. Given the complicated nature of Eq.~\eqref{eq:rr}, we chose to discard this equation and consider only Eqs.~\eqref{eq:tt}, \eqref{eq:Vphicosmo}, \eqref{eq:Vpsicosmo}, and \eqref{eq:conserv-total}.

By introducing the Hubble function $H=\dot{a}/a$, the system of cosmological field equations takes the form
\be\label{Fr1}
3H^2=8\pi \frac{\rho}{\varphi}+\frac{3\psi}{2\varphi}\left(\rho-\frac{1}{3}p\right)+\frac{1}{2}\frac{V}{\varphi}-3H\frac{\dot{\varphi}}{\varphi},
\ee
\be\label{Fr2}
2\dot{H}+3H^2=-8\pi \frac{p}{\varphi}+\frac{\psi}{2\varphi}\left(\rho-3p\right)+\frac{1}{2}\frac{V}{\varphi}-\frac{\ddot{\varphi}}{\varphi}-2H\frac{\dot{\varphi}}{\varphi},
\ee
\be\label{Fr3}
V_{\varphi}=6\left(\dot{H}+2H^2\right), \qquad V_{\psi}=3p-\rho,
\ee
\bea\label{Fr4}
\hspace{-0.3cm}\dot{\rho}+3H(\rho +p)&=&\frac{3}{8\pi}\Bigg\{-\frac{\dot{\psi}}{2}\left(\rho-\frac{p}{3}+\frac{V_{\psi}}{3}\right)\nonumber\\
\hspace{-0.3cm}&&-\psi\left[H(\rho+p)+\frac{1}{2}\left(\dot{\rho}-\frac{1}{3}\dot{p}\right)\right]\Bigg\}.
\eea

As an indicator of the decelerating/accelerating nature of the cosmological evolution we consider the deceleration parameter $q$, defined as
\be
q=\frac{d}{dt}\frac{1}{H}-1=-\frac{\dot{H}}{H^2}-1.
\ee

With the use of the generalized Friedmann equations (\ref{Fr1}) and (\ref{Fr2}) we obtain for $q$ the expression
\be
q=\frac{1}{2}+\frac{3\left[ 4\pi \frac{p}{\varphi }-\frac{\psi }{4\varphi }%
\left( \rho -3p\right) -\frac{V}{4\varphi }+\frac{\ddot{\varphi}}{2\varphi }%
+H\frac{\dot{\varphi}}{\varphi }\right] }{8\pi \frac{\rho }{\varphi }+\frac{%
\psi }{2\varphi }\left( \rho -3p\right) +\frac{V}{2\varphi }-3H\frac{\dot{%
\varphi}}{\varphi }}.
\ee

Note that, in general, modified theories of gravity with geometry-matter couplings \cite{Harko:2011kv,Koivisto:2005yk,Bertolami:2007gv,Harko:2010mv,Harko:2012hm,Haghani:2013oma,Harko:2014gwa,Harko:2014sja,Harko:2014aja,Avelino:2016lpj,Harko:2018gxr,Harko:2020ibn} imply the non-conservation of the matter stress-energy tensor $\nabla_\mu T^{\mu\nu} \neq 0$, which may entail a transfer of energy from the geometry to the matter sector \cite{Harko:2014pqa,Harko:2015pma,Harko:2018ayt,Harko:2021bdi}.

\subsection{Thermodynamic interpretation and matter creation}\label{sec:thermodynamics}

Here, we consider the formalism of irreversible matter creation of thermodynamics of open systems considered in the context of cosmology, as described in the ground breaking work by Prigogine and collaborators \cite{Pri}. In this formalism, the Universe is seen as an open system, and the description of particle creation is based on the reinterpretation of the energy-momentum tensor of matter by including a matter creation term in the conservation laws.

Let us consider an open system of volume $V$ containing $N(t)$ particles, with an energy density $\rho$ and a thermodynamic pressure $p$, respectively. For such a system, the thermodynamical conservation equation, written in its most general form, is given by
\begin{equation}\label{eq:1st law 1}
d(\rho V)=d Q-p d V+\frac{h}{n} d(n V),
\end{equation}
where $dQ$ is the heat received by the system during time $dt$, $h=\rho + p$ is the enthalpy per unit volume and $n=N/V$ is the particle number density. Unlike isolated or closed systems, where the number of particles remain constant, the thermodynamical conservation of energy in open systems contains a term that expresses the matter creation/annihilation processes that can occur within the system. The second law of thermodynamics takes the following form
\begin{equation}
\label{eq:entropy1}
d S=d_{e} S+d_{i} S\geq 0,
\end{equation}
where $d_eS$ is the entropy flow and $d_iS$ is the entropy creation. The first term can be seen as measure of the variation of the system's homogeneity and the latter is the part of entropy that is solely related to matter creation.

To find expressions for these two quantities, one starts by writing the total differential of the entropy
\begin{equation}
\label{eq:total_diff_entropy1}
{\cal T} d S=d(\rho V)+p d V-\mu d(n V),
\end{equation}
where $\mu$ is the chemical potential, and $\cal{T}$ is the thermodynamic temperature. By using Eq. \eqref{eq:1st law 1} and the thermodynamical relation $\mu n=h-{\cal T}s$, where $s=S/V$ is the entropy density,
one can then write Eq. \eqref{eq:total_diff_entropy1} in a more convenient way
\begin{equation}
{\cal T}d S=d Q+{\cal T} \frac{s}{n} d(n V).
\end{equation}
But Eq. \eqref{eq:entropy1} implies that
\begin{equation}
\label{eq:TdS}
{\cal T} d S={\cal T} d_{e} S+{\cal T} d_{i} S,
\end{equation}
and thus it is possible to obtain directly an expression for the entropy flow and for the entropy creation, respectively
\begin{equation}
\label{eq:entropies}
d_{e} S=\frac{d Q}{{\cal T}}, \qquad d_{i} S=\frac{s}{n} d(n V).
\end{equation}

Since entropy flow can be seen as a measure of the variation of the system's homogeneity, if we consider an homogeneous system, the variation of homogeneity is zero. Therefore, the entropy flow vanishes in a system with this specific configuration. This means that in homogeneous systems we expect adiabatic processes to occur and for that reason matter creation is the only contribution to entropy production
\begin{equation}
\label{Entropy3}
d S=d_{i} S=\frac{s}{n} d(n V)\geq0.
\end{equation}

We now apply the formalism of irreversible matter creation of thermodynamics of open systems to cosmology. For that, let us consider an homogeneous and isotropic Universe as an open system with volume $V$ containing $N(t)$ particles, an energy density $\rho$ and thermodynamic pressure $p$, described by Eq. \eqref{eq:FLRW-metric}.
Since we are considering an isotropic and homogeneous Universe, the volume can be expressed in terms of the scale factor, $V=a^3(t)$. Then, Eq. \eqref{eq:1st law 1} can be rewritten in terms of (total) time derivatives of the physical quantities as
\begin{equation}\label{eq:1st law 2}
\frac{d}{d t}\left(\rho a^{3}\right)+p \frac{d}{d t} a^{3}=\frac{d Q}{d t}+\frac{\rho+p}{n} \frac{d}{d t}\left(n a^{3}\right).
\end{equation}
As we have seen before, homogeneous systems do not receive heat. Since the Universe we are considering is homogeneous, we conclude that the heat received by it remains constant over time, i.e $dQ/dt=0$. This result allow us to reformulate Eq. \eqref{eq:1st law 2} in an equivalent form
\begin{equation}\label{eq:1st law 3}
\dot{\rho}+3 H(\rho+p)=\frac{\rho+p}{n}(\dot{n}+3 H n).
\end{equation}

Furthermore, by comparing Eq. \eqref{eq:1st law 3} with Eq. \eqref{eq:conserv-total} we conclude that, in the formalism of thermodynamics of open systems, the latter also constitutes the thermodynamical conservation equation for the Universe, in which the presence of the two scalar fields, $\varphi$ and $\psi$, contribute to particle creation in an homogeneous and isotropic geometry, with the time variation of the particle number density obtained as
\begin{equation}\label{eq: number density variation}
    \dot{n}+3 H n = \Gamma n,
\end{equation}
where $\Gamma$ is the particle creation rate. Substituting Eq. \eqref{eq: number density variation} into Eq. \eqref{eq:1st law 3} we get the energy conservation equation in an alternative form
\begin{equation}\label{eq:HGamma}
\dot{\rho}+3 H(\rho+p)=(\rho+p)\Gamma.
\end{equation}

Therefore, by using Eq. \eqref{eq:conserv-total} and Eq. \eqref{eq:HGamma} we find that the particle creation rate in the scalar-tensor representation of $f(R,T)$ gravity assumes the following form
\begin{eqnarray}\label{eq:gamma}
\Gamma &=&\frac{3}{8 \pi} \frac{1}{\rho+p} \left[\dot{\varphi}\left(\frac{\ddot{a}}{a}+\frac{\dot{a}^{2}}{a^{2}}-\frac{1}{6} V_{\varphi}\right)   \right.
	\nonumber  \\
&& \qquad -\dot{\psi}\left(\frac{1}{2}\rho-\frac{1}{6} p+\frac{1}{6} V_{\psi}\right)
	\nonumber \\
&& \qquad \left.-\psi\left(\frac{\dot{a}}{a}(\rho+p)+\frac{1}{2} \dot{\rho}-\frac{1}{6} \dot{p}\right)\right].
\end{eqnarray}

Substituting $V_{\varphi}$ and $V_{\psi}$ by their expressions, Eqs. \eqref{eq:Vphicosmo} and \eqref{eq:Vpsicosmo}, respectively, and using Eq. \eqref{eq:HGamma} we obtain a simplified expression for the particle creation rate
\begin{equation}\label{eq:gamma 2}
\Gamma=-\frac{\psi}{8 \pi+\psi}\left(\frac{d}{d t} \ln \psi+\frac{1}{2} \frac{\dot{\rho}-\dot{p}}{\rho+p}\right).
\end{equation}

For adiabatic transformations describing irreversible particle creation
in an open thermodynamic system, the first law of thermodynamics can be rewritten as an effective energy conservation equation
\begin{equation}
\frac{d}{d t}\left(\rho a^{3}\right)+\left(p+p_{c}\right) \frac{d}{d t} a^{3}=0,
\end{equation}
where we have introduced the creation pressure $p_c$, a supplementary pressure that must be considered in open systems because of irreversible matter creation processes. Expressing the equation above in an equivalent manner
\begin{equation}
\label{eq:effective_energy_conservation}
\frac{d}{d t}\left(\rho a^{3}\right)+p \frac{d}{d t} a^{3}=-p_{c}\frac{d}{d t}a^3
\end{equation}
and comparing it with Eq. \eqref{eq:1st law 2}, after some simplifications we can write the creation pressure as
\begin{equation}
\label{eq: creation pressure1}
p_{c}=-\frac{\rho+p}{3 H} \; \Gamma.
\end{equation}

Therefore, to determine the creation pressure it is enough to know the particle creation rate. Consequently, by using Eq. \eqref{eq:gamma 2}, we find that the creation pressure in the scalar-tensor representation of $f(R,T)$ gravity takes the following form
\begin{equation}\label{eq: creation pressure 2}
p_{c}=\frac{\rho+p}{3 H} \frac{\psi}{8 \pi+\psi}\left(\frac{d}{d t} \ln \psi+\frac{1}{2} \frac{\dot{\rho}-\dot{p}}{\rho+p}\right).
\end{equation}
These results are consistent with \cite{Harko:2014pqa} because of our definition of the field $\psi$, present in Eq. \eqref{eq:varphi&psi}, therefore proving the equivalence between the geometrical representation and the scalar-tensor representation of $f(R,T)$ gravity.

Note that both the creation rate and creation pressure only depend on the scalar field $\psi$, which is the one associated with the trace of the energy-momentum tensor. This means there is a contribution to matter creation from the variation in the degree of freedom that encapsulates the coupling between geometry and matter. Hence, we conclude that geometry-matter couplings induce particle production. In the limit where the function $f$ does not depend on $T$, we regain the usual $f(R)$ gravity theory in which the covariant divergence of the energy-momentum tensor of matter is zero. This conservation, as it happens with GR, lead to the vanishing of the creation rate and of the creation pressure.

\subsection{Entropy and temperature evolution}

We now recall the 2nd law of thermodynamics in the context of open systems, Eq. \eqref{eq:entropy1}, to explore the entropy evolution. We saw earlier that the condition of homogeneity implies the vanishing of the entropy flow, $d_eS=0$, which means the only contribution to entropy production is due to entropy creation, and consequently Eq. \eqref{eq:entropy1} becomes Eq. \eqref{Entropy3}. Following our previous assumptions, that the Universe is both homogeneous and isotropic, by taking the total time derivative of Eq. \eqref{Entropy3} and using the expression for the entropy creation, present in Eq. \eqref{eq:entropies}, the comoving volume written in terms of the scale factor, $V=a^3(t)$, the definition of entropy density ($s=S/V$) and Eq. \eqref{eq: number density variation}, one can obtain the following expression for the entropy temporal evolution
\begin{equation}
\label{eq:entropy_evol}
\frac{d S}{d t}=\Gamma S  \geq 0,
\end{equation}
whose general solution is
\begin{equation}
\label{S(t)}
S(t)=S_{0} \text{exp}\left[\int_{0}^{t} \Gamma\left(t^{\prime}\right) d t^{\prime}\right],
\end{equation}
with $S_0=S(0)$ constant. Therefore, in an homogeneous and isotropic geometry, in the formalism of irreversible matter creation, what causes the time variation of the entropy is the particle creation rate. Substituting Eq. \eqref{eq:gamma 2} into Eq. \eqref{S(t)}, we obtain the expression for the entropy in scalar-tensor $f(R,T)$ gravity
\begin{equation}
S(t)=S_{0} \text{exp}\left[-\int_{0}^{t} \frac{\psi}{8 \pi+\psi}\left(\frac{d}{d t} \ln \psi+\frac{1}{2} \frac{\dot{\rho}-\dot{p}}{\rho+p}\right) d t\right].
\end{equation}

The entropy flux 4-vector $S^{\mu}$ is defined as \cite{Calvao:1991wg}
\begin{equation}
S^{\mu}=n \sigma u^{\mu} \,,
\end{equation}
where $\sigma=S/N$ is the entropy per particle (or characteristic entropy). Since $S^{\mu}$ must obey the 2nd law of thermodynamics, hence we have the following condition
\begin{equation}
    \nabla_{\mu} S^{\mu}\geq0,
\end{equation}
which is the second law of thermodynamics written in a covariant formulation. Therefore, to obtain the entropy production rate due to matter creation processes, one must determine the covariant derivative of the entropy flux 4-vector
\begin{equation}
\nabla_{\mu} S^{\mu} = \left(\nabla_{\mu}n\right) \sigma u^{\mu}+n\left(\nabla_{\mu} \sigma \right) u^{\mu}+n\sigma \nabla_{\mu}u^{\mu},
\end{equation}
which by using $\nabla_{\mu}u^{\mu}=3H$ and $u^{\mu}\nabla_{\mu}=d/dt$ assumes the following form
\begin{equation}
\label{entropy_prod_rate1}
\nabla_{\mu} S^{\mu}=(\dot{n}+3 H n)\sigma + n\dot{\sigma}.
\end{equation}

To further simplify the expression above we take the time derivative of the Gibbs relation \cite{Calvao:1991wg}
\begin{equation}
n {\cal T} \dot{\sigma}= \dot{\rho}-\frac{\rho + p}{n} \dot{n},
\end{equation}
and use it in combination with the expression for the chemical potential
\begin{equation}
\mu=\frac{h}{n}-{\cal T}\frac{s}{n}=\frac{\rho + p}{n}-{\cal T} \sigma ,
\end{equation}
alongside Eqs.~\eqref{eq:1st law 3} and \eqref{eq: number density variation}. With that, we obtain a compact form for the covariant derivative of the entropy flux 4-vector
\begin{equation}
\label{entropy_prod_rate}
    \nabla_{\mu} S^{\mu} =\Gamma s \geq 0.
\end{equation}

One can explore the similarities between Eqs.~\eqref{eq:entropy_evol} and \eqref{entropy_prod_rate}. Both entropy temporal evolution and entropy production rate depend on the particle creation rate, evidencing the fundamental role played by this quantity in the description of an homogeneous and isotropic Universe, in which matter creation processes occur. The only difference between the two is that the entropy production rate depends on the entropy density (as expected since we have a flux), while the entropy temporal evolution depends on the entropy itself. Substituting Eq. \eqref{eq:gamma 2} into Eq. \eqref{entropy_prod_rate} we finally obtain the entropy production rate in the scalar-tensor $f(R,T)$ gravity
\begin{equation}
\nabla_{\mu} S^{\mu}=-\frac{\psi}{8 \pi+\psi}\left(\frac{d}{d t} \ln \psi+\frac{1}{2} \frac{\dot{\rho}-\dot{p}}{\rho+p}\right)s \geq 0.
\end{equation}

Our objective now is to obtain the temperature evolution, similar to what was done above. A thermodynamic system is fundamentally described by the particle number density $n$ and the temperature $\cal{T}$. In a thermodynamic equilibrium situation, the energy density $\rho$ and the pressure $p$ are determined from the equilibrium equations of state, given in a parametric form as
\begin{equation}
\label{drho_and_dp}
\rho=\rho(n, {\cal T}), \qquad
p=p(n, {\cal T}).
\end{equation}
Then, the differential of the energy density and the differential of the pressure are, respectively
\begin{equation}
\label{drho}
d \rho=\left(\frac{\partial \rho}{\partial n}\right)_{{\cal T}} d n+\left(\frac{\partial \rho}{\partial {\cal T}}\right)_{n} d {\cal T},
\end{equation}
\begin{equation}
d p=\left(\frac{\partial p}{\partial n}\right)_{{\cal T}} d n+\left(\frac{\partial p}{\partial {\cal T}}\right)_{n} d{\cal T},
\end{equation}
where the subscripts $\cal{T}$ and $n$ on the partial derivatives indicate that the temperature $\cal{T}$ and the particle number $n$ are fixed, respectively.
Recalling the energy conservation equation obtained previously, Eq.~\eqref{eq:HGamma}, and inserting Eq.~\eqref{drho} in it, we obtain
\begin{equation}
\label{energy_conserv_temp}
\left(\frac{\partial \rho}{\partial n}\right)_{{\cal T}} \dot{n}+\left(\frac{\partial \rho}{\partial {\cal T}}\right)_n \dot{{\cal T}}+3(\rho+p) H=(\rho+p) \Gamma.
\end{equation}

To express the energy conservation equation above in a more convenient manner, first we use the Gibbs relation \cite{Calvao:1991wg} to write the differential of the characteristic entropy $\sigma$ as
\begin{equation}
\label{dsigma1}
d\sigma=\frac{1}{n {\cal T}} d \rho-\frac{\rho+p}{n^{2} {\cal T}} dn.
\end{equation}
By looking at Eq.~\eqref{dsigma1}, one could say that $\sigma$ is a function of $\rho$ and $n$. However, since $\rho$ itself is a function of $n$ and $\cal{T}$ (Eq. \eqref{drho_and_dp}), thus $\sigma$ is, fundamentally, a function of $n$ and $\cal{T}$. By this reasoning, the true differential of the characteristic entropy is
\begin{equation}
\label{dsigma_2}
d \sigma=\left(\frac{\partial \sigma}{\partial n}\right)_{{\cal T}} d n+\left(\frac{\partial \sigma}{\partial {\cal T}}\right)_{n} d {\cal T}.
\end{equation}
To obtain an explicit expression for this differential, one just inserts Eq.~\eqref{drho} into Eq.~\eqref{dsigma1}, yielding
\begin{equation}
    d\sigma=\left[\frac{1}{n{\cal T}}\left(\frac{\partial \rho}{\partial n}\right)_{{\cal T}}+\frac{\rho + p}{n^2{\cal T}}\right] dn + \frac{1}{n{\cal T}}\left(\frac{\partial \rho}{\partial {\cal T}}\right)_{n}d{\cal T}.
\end{equation}
The entropy $S$ is an exact differential, and so it is the characteristic entropy $\sigma$. Therefore the condition,
\begin{equation}
    \left[\frac{\partial}{\partial {\cal T}}\left(\frac{\partial \sigma}{\partial n}\right)_{{\cal T}}\right]_{n}=\left[\frac{\partial}{\partial n} \left(\frac{\partial \sigma}{\partial {\cal T}}\right)_{n}\right]_{{\cal T}},
\end{equation}
follows immediately. Then, one can obtain the following thermodynamical relation
\begin{equation}
\left(\frac{\partial \rho}{\partial n}\right)_{{\cal T}}=\frac{\rho+p}{n}-\frac{{\cal T}}{n}\left(\frac{\partial p}{\partial {\cal T}}\right)_{n},
\end{equation}
which is plugged into the energy conservation equation \eqref{energy_conserv_temp}, and with the help of Eqs.~\eqref{eq:1st law 3} and \eqref{eq:HGamma} we achieve an expression for the temperature evolution
\begin{equation}
\frac{1}{{\cal T}} \frac{d {\cal T}}{d t}=c_{s}^{2} \frac{\dot{n}}{n},
\end{equation}
where $c_s=\sqrt{\left(\partial p/ \partial \rho\right)_{n}}$ is the speed of sound. By using Eq.~\eqref{eq: number density variation}, we write the temperature evolution in terms of the particle creation rate as the following equation,
\begin{equation}
\frac{1}{{\cal T}} \frac{d {\cal T}}{d t}=c_{s}^{2}(\Gamma-3 H),
\end{equation}
whose solution is
\begin{equation}
{\cal T}(t)={\cal T}_{0} \text{exp}\left\{c_s^2 \int_{0}^{t^{\prime}}\left[ \Gamma\left(t^{\prime}\right)-3 H\left(t^{\prime}\right)\right] d t^{\prime}\right\},
\end{equation}
where ${\cal T}_0={\cal T}(0)$ is a constant. Because of the complicated for of the Hubble function $H$ in the present model, a solution for the temperature should be obtained numerically. Finally, the temperature as a function of time in scalar-tensor $f(R,T)$ gravity is given by
\begin{eqnarray}
&{\cal T}(t)={\cal T}_{0} \text{exp}\left\{c_s^2 \int_{0}^{t} \left[\frac{\psi}{8 \pi+\psi}\left(\frac{d}{d t} \ln \psi+\frac{1}{2} \frac{\dot{\rho}-\dot{p}}{\rho+p}\right) -3H\right] d t\right\}.
\nonumber \\
\end{eqnarray}

\section{Particular cosmological models}\label{sec:models}

In the present Section, we will consider several specific cosmological models, obtained by specifying in advance the functional form of the potential $V(\varphi,\psi)$. The existence of the de Sitter solution will also be investigated. We will also consider a detailed comparison of the theoretical models  with cosmological observations.

The comparison between different theoretical models and observations  can be performed in a much easier way if one introduces as independent variable the
redshift $z$, defined as $1 + z =1/a$. Then
\be
\frac{d}{dt}=\frac{dz}{dt}\frac{d}{dz}=-(1+z)H(z)\frac{d}{dz}.
\ee

In the $\Lambda$CDM model the energy density of dust  matter with $p=0$ scales according to the law  $\rho= \rho
_0/a^3=\rho _0 (1+z)^3$, where $\rho _0$ is the present day matter density. This relation directly follows from the law of the conservation of the energy-momentum tensor.
As a function of the scale factor the time evolution of the Hubble function is given in the $\Lambda$CDM model, as a function of the
scale factor, by \cite{Harko:2018ayt}
\begin{equation}
H=H_0\sqrt{\left(\Omega _b+\Omega _{DM}\right)a^{-3}+\Omega _{\Lambda}},
\end{equation}
where $H_0$ is the present day value of the Hubble function, while by $\Omega _b$, $\Omega _{DM}$, and $\Omega _{\Lambda}$ we have denoted
the density parameters of the baryonic matter, of the cold (pressureless)
dark matter, and of the dark energy (modeled by a cosmological constant),
respectively. The three density parameters satisfy the simple algebraic relation $\Omega _b+\Omega
_{DM}+\Omega _{\Lambda}=1$, which follows from the flatness of the geometry of the Universe.

In the standard $\Lambda$CDM  model the deceleration parameter is obtained as
\begin{equation}
q(z)=\frac{3 (1+z)^3 \left(\Omega _{DM}+\Omega _b\right)}{2 \left[\Omega
_{\Lambda}+(1+z)^3 \left(\Omega _{DM}+\Omega _b\right)\right]}-1.
\end{equation}

In the following for the matter density parameters we adopt the values \cite{1h,1h1}, with $\Omega
_{DM}=0.2589$, $\Omega _{b}=0.0486$, and $\Omega _{\Lambda}=0.6911$%
, respectively, obtained from the CMB spectrum as investigated by the Planck satellite.  The total matter
density parameter $\Omega _m=\Omega _{DM}+ \Omega _b$ is then given by $\Omega _m=0.3089$. The present day value of the deceleration parameter is $q(0)=-0.5381$,
which indicates that present day Universe is in an accelerating state.

\subsection{The de Sitter solution}

The de Sitter solution corresponds to a constant Hubble function, $H=H_0={\rm constant}$. We now investigate the possibility of the existence of the de Sitter solution in the two scalar field-tensor representation of $f(R,T)$ gravity.

\subsubsection{The constant density solution}

We assume that matter consists of pressureless dust with $p=0$, and that during the exponentially expanding phase the matter density is constant, $\rho=\rho_0={\rm constant}$. Then Eqs.~(\ref{Fr3}) can be easily integrated to give
\be
V(\varphi,\psi)=12H_0^2\varphi +f(\psi),
\ee
and
\be
V(\varphi, \psi)=-\rho_0 \psi +g(\varphi),
\ee
where $f(\psi)$ and $g(\varphi)$ are arbitrary functions of the argument, giving
\be
V(\varphi, \psi)=12H_0^2\varphi -\rho _0 \psi+\Lambda _0,
\ee
where $\Lambda _0$ is an arbitrary constant of integration. From Eq.~(\ref{Fr4}) we obtain for $\psi$ the evolution equation
\be
\dot{\psi}+3H_0\psi=24\pi H_0,
\ee
with the general solution given by
\be
\psi (t)=e^{-3H_0t}\left[\psi _0-8\pi\left(1- e^{3H_0t}\right)\right],
\ee
where $\psi _0=\psi(0)$. In the large time limit $\psi(t)$ tends to a constant, $\lim_{t\rightarrow \infty}\psi(t)=8\pi$. Then Eq.~(\ref{Fr1}) gives the evolution equation for $\varphi$,
\be
\dot{\varphi}-H_0\varphi =\frac{\Lambda _0}{6H_0}+\frac{16\pi \rho_0}{3H_0}-\frac{e^{-3H_0t}}{\rho_0\left(8\pi -\psi_0\right)},
\ee
with the general solution given by
\bea
\varphi (t)&=&\frac{1}{12H_{0}^{2}}\Big\{\Big[
e^{H_{0}t}\left( 12H_{0}^{2}\varphi _{0}+2\Lambda _{0}+%
\rho _{0}\psi _{0}  +56\pi \rho _{0}\right)
	\nonumber\\
&&-2(\Lambda _{0}+32\pi \rho _{0})-\rho _{0} e^{-3H_{0}t} \left(
8\pi -\psi _{0}\right) \Big] \Big\},
\eea
where $\varphi _0=\varphi (0)$. The constant density of the Universe is maintained by matter creation with particle production rate
\be
\Gamma =\frac{3 H_0 (\psi_0-8 \pi)}{8 \pi
   \left(2 e^{3 H_0
   t}-1\right)+\psi_
   0}.
\ee

The corresponding creation pressure can be written as
\be
p_c=-\frac{\rho_0}{3H_0}\Gamma=-\frac{\rho_0 (\psi_0-8 \pi)}{8 \pi
   \left(2 e^{3 H_0
   t}-1\right)+\psi_
   0}.
\ee

\subsubsection{de Sitter evolution with time varying matter density}

A de Sitter type solution $H=H_0={\rm constant}$ in the two scalar field representation of $f(R,T)$ gravity can also be obtained for a non-constant matter density, with matter in the form of dust with $p=0$. To show this we consider that the potential $V$ has the form
\be
V(\varphi, \psi)=12H_0^2\varphi-\frac{1}{2\beta} \psi ^2,
\ee
where $\beta$ is a constant. The first of the Eqs.~(\ref{Fr3}) is then identically satisfied, while the second gives
\be
\psi =\beta \rho.
\ee
Thus the potential becomes
\be
V(\varphi,\psi)=12H_0^2\varphi-\frac{\beta }{2}\rho ^2.
\ee

Equation~(\ref{Fr4}) can then be written as a first order nonlinear differential equation for $\rho$,
\be
\left(1+\frac{5\beta}{16\pi }\rho\right)\dot{\rho}+3H_0\rho=-\frac{3\beta H_0}{8\pi}\rho^2,
\ee
with the general solution given by
\be\label{55}
\rho \left(\beta \rho +8\pi\right)^{3/2}=e^{-3H_0\left(t-t_0\right)},
\ee
where $t_0$ is an arbitrary constant of integration. By substituting the expression of $V$ in Eq.~(\ref{Fr1}), and by taking into account that
\begin{equation}
\frac{d\varphi }{dt}=\frac{d\varphi }{d\rho }\frac{d\rho }{dt}=-\frac{%
3H_{0}\rho \left[ 1+\left( \beta /8\pi \right) \rho \right] }{1+\left(
5\beta /16\pi \right) \rho }\frac{d\varphi }{d\rho },
\end{equation}
Eq.~(\ref{Fr1}) becomes
\begin{equation}
9H_{0}^{2}\frac{\rho \left[ 1+\left( \beta /8\pi \right) \rho \right] }{%
1+\left( 5\beta /16\pi \right) \rho }\frac{d\varphi }{d\rho }%
+3H_{0}^{2}\varphi +8\pi \rho +\frac{5\beta }{4}\rho ^{2}=0,
\end{equation}
with the general solution given by
\bea
\varphi (\rho )&=&-\frac{1}{11220\beta H_{0}^{2}}\Bigg[
\frac{55\beta }{\sqrt[3]{\rho }} \left( 25\beta \rho ^{7/3}-\frac{204c_{1}H_{0}^{2}}{\sqrt{%
\beta \rho +8\pi }}\right)
	\nonumber\\
&&  \qquad -\frac{27648\pi
^{5/2}\, }{\sqrt{\frac{\beta \rho }{2}+4\pi }}\,
_{2}F_{1}\left( \frac{1}{3},\frac{1}{2};\frac{4}{3};-\frac{\beta
\rho }{8\pi }\right)
	\nonumber \\
&& \qquad  +13824\pi ^{2} +6400\pi \beta \rho
\Big],
\eea
where $c_1$ is an arbitrary constant of integration, and  $_2F_1(a,b;c;z)$ is the hypergeometric function $_2F_1(a,b;c;z)=\sum _{k=0}^{\infty}{\left((a)_k(b)_k/(c)_k\right)\left(z^k/k!\right)}$. Finally, the particle creation rate can be obtained as
\begin{eqnarray}
\Gamma &=& -\frac{3}{2}\frac{\beta \rho }{\beta \rho +8\pi }\frac{\dot{\rho}}{%
\rho }
	\nonumber \\
&=&\frac{9H_{0}\beta }{2}\frac{\rho \left[ 1+\left( \beta /8\pi \right)
\rho \right] }{\left( \beta \rho +8\pi \right) \left[ 1+\left( 5\beta /16\pi
\right) \rho \right] }.
\end{eqnarray}

Hence, the general solution of the generalized Friedmann equations describing a de Sitter type expansion can be obtained in an exact parametric form, with $\rho$ taken as parameter. The time variation of the density can be easily obtained in the limits $\beta \rho \ll 8\pi$ and $\beta \rho \gg  8\pi$, respectively, as $\rho(t)\sim e^{-3H_0\left(t-t_0\right)}$ and $\rho (t)\sim e^{-(6/5)H_0\left(t-t_0\right)}$, respectively. For $\beta \rho \gg 8\pi$, the matter creation rate becomes a constant, $\Gamma \approx (9/5)H_0$, while in the opposite limit $\Gamma \approx \left(9H_0/16\pi\right)e^{-3H_0\left(t-t_0\right)}$.

\subsection{$V(\varphi,\psi)=\alpha \varphi +\beta \varphi ^2-\left(1/2\gamma\right) \psi ^2$}

We consider now more general cosmological models in the two scalar field representation of $f(R,T)$ gravity, by assuming for the potential $V$ a simple additive quadratic form, namely, $V(\varphi,\psi)=\alpha \varphi +\beta \varphi ^2-\left(1/2\gamma\right) \psi ^2$, where $\alpha$, $\beta$  and $\gamma$ are constants. We will consider only the case of a dust Universe, and hence we take $p=0$. Then the second of Eqs.~(\ref{Fr3}) gives $\psi =\gamma \rho$. Hence the system of the cosmological evolution equations can be formulated as the following first order dynamical system,
\be\label{a1}
\dot{H}=\frac{\alpha}{6}+\frac{\beta }{3}\varphi -2H^2,
\ee
\begin{equation}\label{a2}
\dot{\rho}=-3H\frac{\rho \left[ 1+\left( \gamma /8\pi \right) \rho \right] }{%
1+\left( 5\gamma /16\pi \right) \rho },
\end{equation}
and
\begin{equation}\label{a3}
\dot{\varphi}=\frac{8\pi }{3}\frac{\rho }{H}+\frac{5\gamma }{12}\frac{\rho
^{2}}{H}+\frac{\alpha }{6}\frac{\varphi}{H} +\frac{\beta }{6}\frac{\varphi ^{2}}{H}-H\varphi ,
\end{equation}
respectively.

In order to simplify the mathematical formalism we introduce a set of dimensionless and rescaled variables $\left(h,\tau,r,\sigma, \epsilon\right)$, defined as
\begin{eqnarray*}
H&=&H_0h,\qquad  \tau =H_0t, \qquad\rho =\frac{3H_0^2 }{8\pi}r, \nonumber\\
\alpha &=&6H_0^2\xi, \;\;\;\;\;  \beta =3H_0^2\sigma,  \;\;\;\; \gamma= \frac{64\pi^2}{3H_0^2} \epsilon,
\end{eqnarray*}
respectively, where $H_0$ is the present day value of the Hubble function. In the new variables the field equations (\ref{a1})-(\ref{a3}) take the form
\be\label{65}
\frac{dh}{d\tau}=\xi+\sigma \varphi-2h^2,
\ee
\be\label{66}
\frac{dr}{d\tau}=-3h\frac{r(1+\epsilon r)}{1+(5\epsilon/2)r},
\ee
\be\label{67}
\frac{d\varphi }{d\tau}=\frac{r}{h}+\frac{5}{4}\epsilon \frac{r^2}{h}+\xi \frac{\varphi}{h}+\frac{\sigma}{2} \frac{\varphi ^2}{h}-h\varphi.
\ee

In the reshift space Eqs.~(\ref{65})-(\ref{67}) are formulated as
\be\label{m1}
-(1+z)h(z)\frac{dh(z)}{dz}=\xi+\sigma \varphi (z) -2h^2(z),
\ee
\be\label{m2}
-(1+z)h(z)\frac{dr(z)}{dz}=-3h(z)\frac{r(z)(1+\epsilon r(z))}{1+(5\epsilon/2)r(z)},
\ee
and
\bea\label{m3}
-(1+z)h(z)\frac{d\varphi (z)}{dz}&=&\frac{r(z)}{h(z)}+\frac{5\epsilon}{4} \frac{r^2(z)}{h(z)}+\xi \frac{\varphi (z)}{h(z)}\nonumber\\
&&+\frac{\sigma}{2} \frac{\varphi ^2(z)}{h(z)}-h(z)\varphi (z).
\eea

\begin{figure*}[htbp]
	\centering
	\includegraphics[scale=0.65]{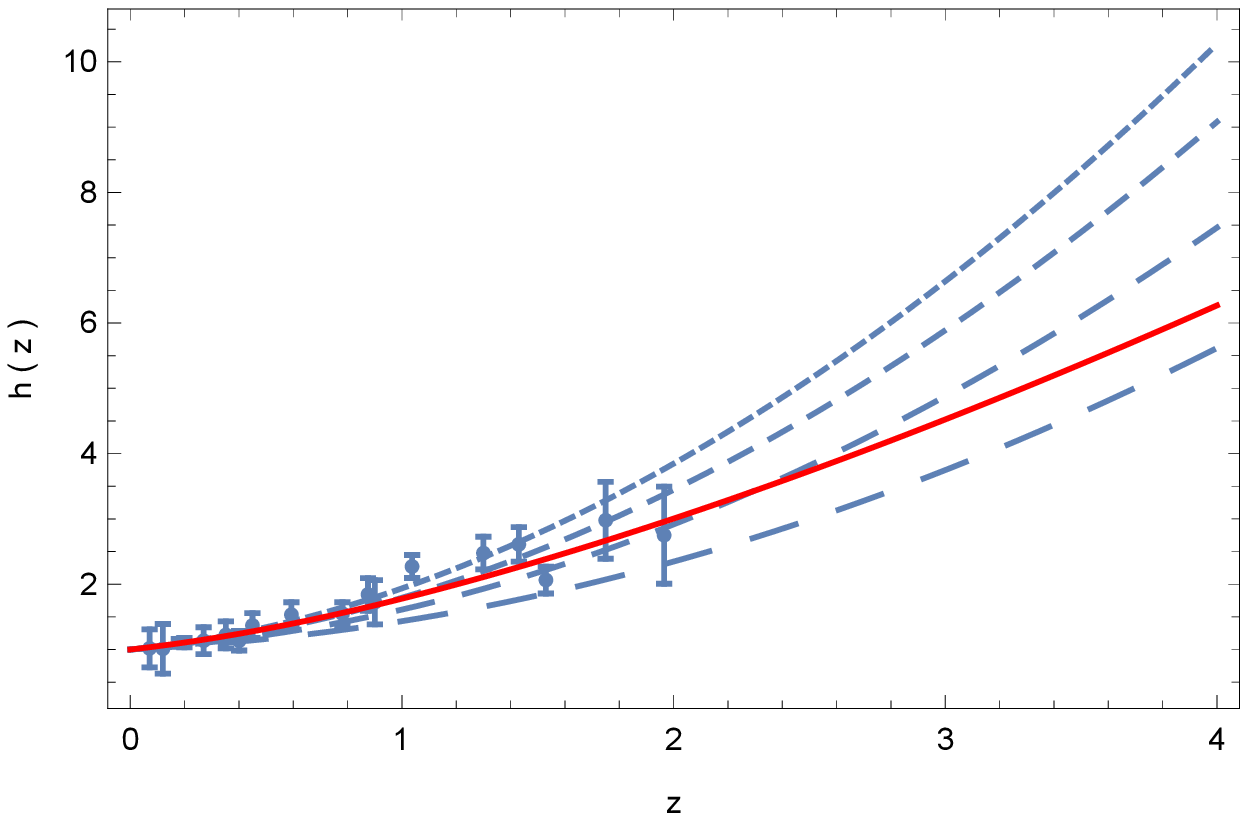}
\includegraphics[scale=0.65]{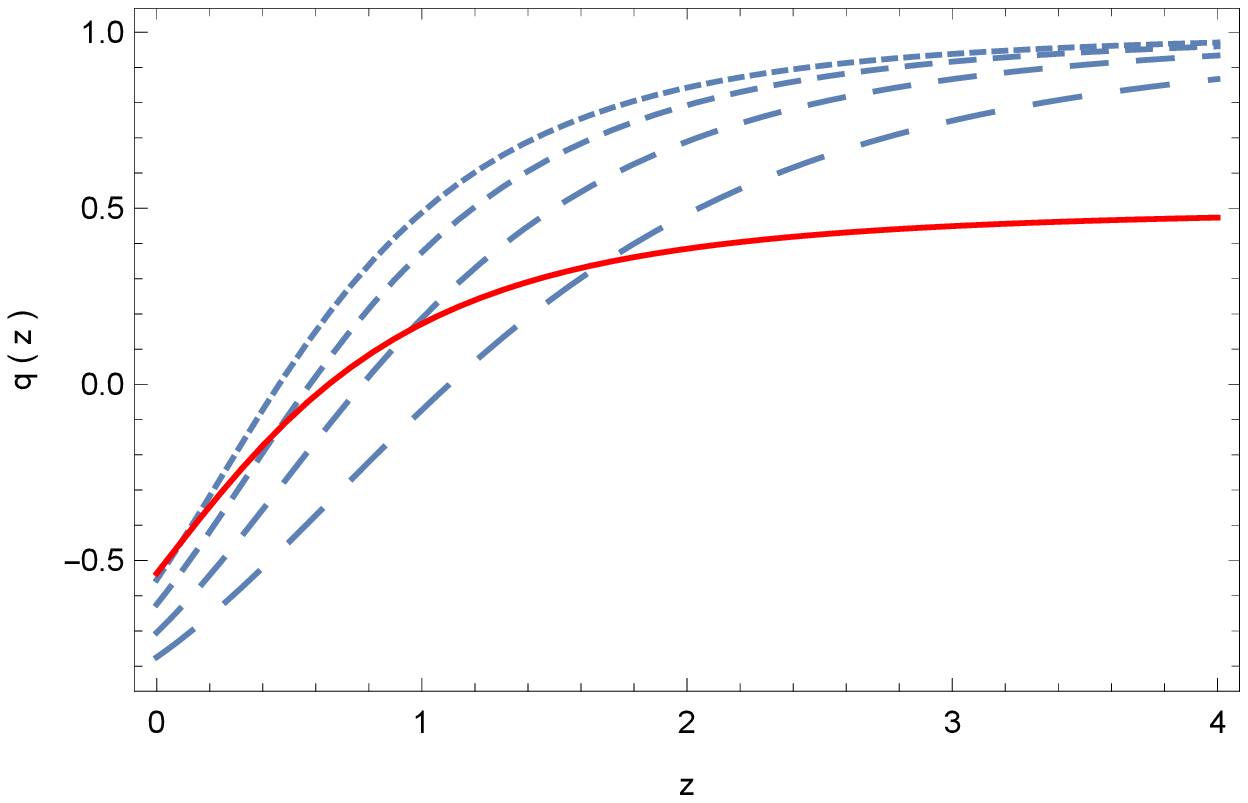}
	\caption{The evolution of the Hubble function $h(z)$  (left panel), and of the deceleration parameter $q(z)$ (right panel)  as a function of redshift $z$ for $\xi=1.54$ (short dashed curve), $\xi=1.61$ (dashed curve), $\xi =1.69$ (long dashed curve), and $\xi=1.76$ (ultra long dashed curve). The values of the potential parameters $\sigma$ and $\epsilon$ have been fixed to $\sigma =-2.89$ and $\epsilon=0.015$, respectively. The initial conditions used for the integration of the cosmological equations are $h(0)=1$, $r(0)=1$ and $\varphi (0)=-0.005$.  The solid red line indicates the predictions of the $\Lambda$CDM model.  The error bars in the left panel indicate the observational values of the Hubble function \cite{H1,H2}.}\label{fig1}
\end{figure*}

\begin{figure*}[htbp]
	\centering
	\includegraphics[scale=0.65]{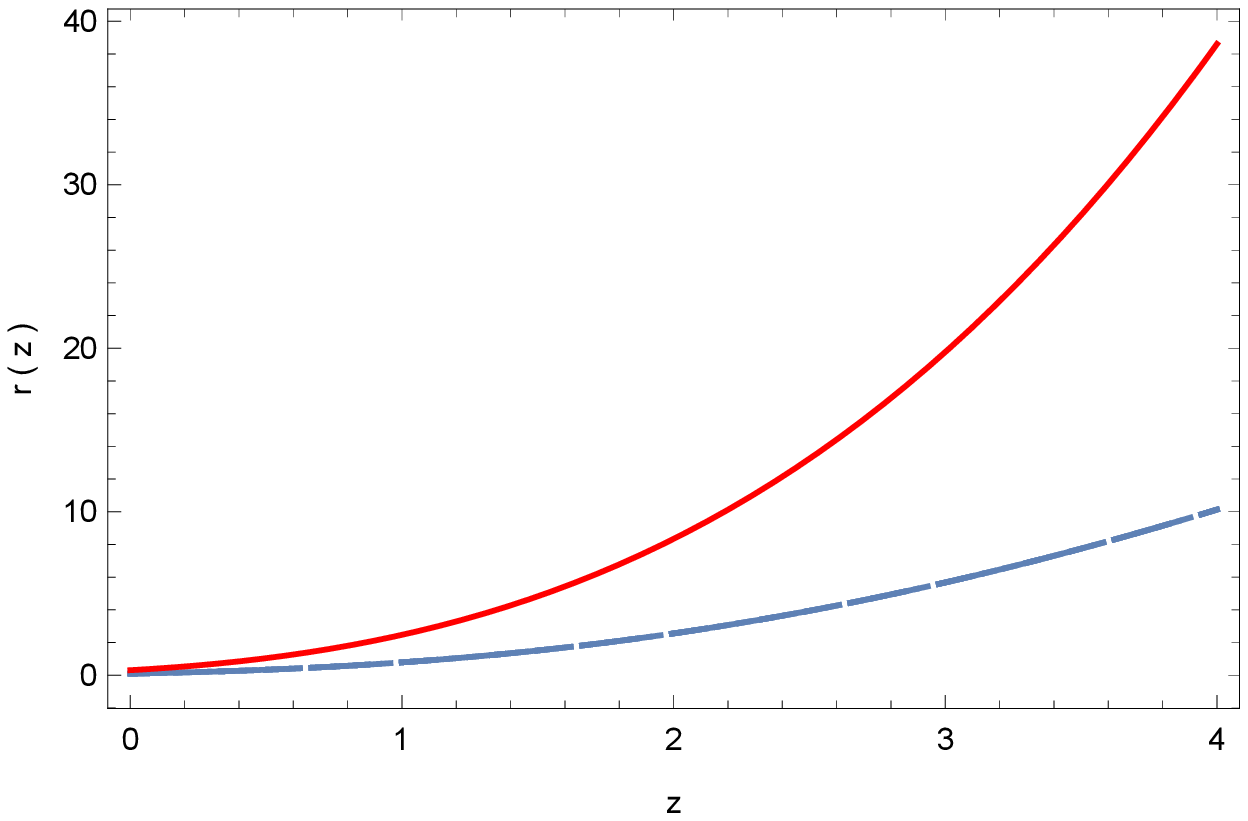}
\includegraphics[scale=0.65]{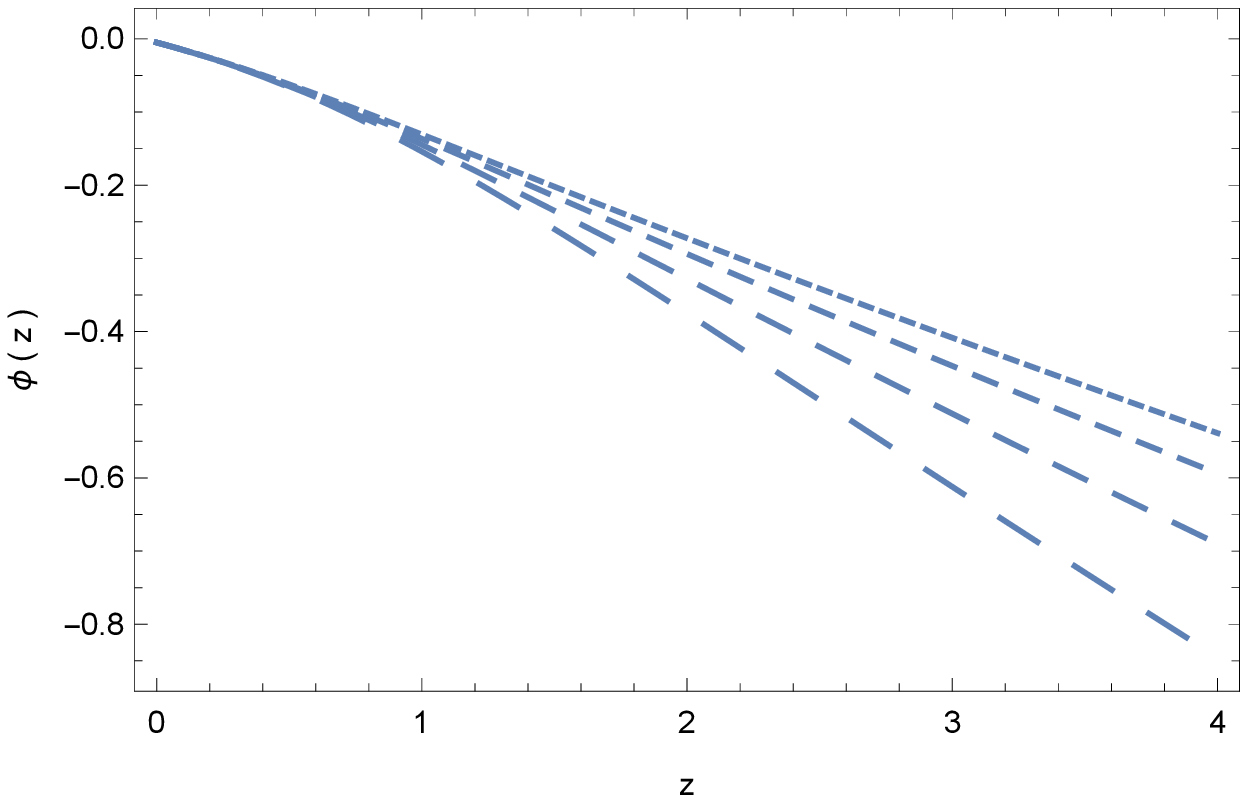}
	\caption{The evolution of the matter energy density $r(z)$  (left panel), and of the function $\varphi(z)$ (right panel)  as a function of redshift $z$ for $\xi=1.54$ (short dashed curve), $\xi=1.61$ (dashed curve), $\xi =1.69$ (long dashed curve), and $\xi=1.76$ (ultra long dashed curve). The values of the potential parameters $\sigma$ and $\epsilon$ have been fixed to $\sigma =-2.89$ and $\epsilon=0.015$, respectively. The initial conditions used for the integration of the cosmological equations are $h(0)=1$, $r(0)=1$ and $\varphi (0)=-0.005$.  The solid red line indicates the predictions of the $\Lambda$CDM model.  }\label{fig2}
\end{figure*}

The results of the numerical integration of the system (\ref{m1})-(\ref{m3}) are presented in Figs.~\ref{fig1} and \ref{fig2}, respectively, for fixed values of the potential parameters $\sigma$ and $\epsilon$, and of the initial condition for $\varphi$, and for different values of the parameter $\xi$. A comparison with the observational data for the Hubble function \cite{H1,H2} and with the predictions of the standard $\Lambda$CDM model is also performed.

The variation of the Hubble function is shown in the left panel of Fig.~\ref{fig1}. The Hubble function is a monotonically increasing function of $z$ (a monotonically decreasing function of time). Its evolution is strongly dependent on the numerical values of the parameters of the potential $V$, as well as on the numerical value of the initial condition for $\varphi (z)$. For certain parameter values the two scalar field representation of $f(R,T)$ gravity can give a good description of the observational data, and can reproduce the predictions of the $\Lambda$CDM model. On the other hand, for higher redshifts $z\geq 3$, significant deviations from the predictions of $\Lambda$CDM appear. Important deviations as compared to standard cosmology are also present in the behavior of the deceleration parameter, shown in the right panel of Fig.~\ref{fig1}. At higher redshifts the considered models have higher values of $q$, indicating a rate of deceleration higher than in $\Lambda$CDM. On the other hand, at lower redshifts the deceleration parameter indicates a higher acceleration rate than in standard cosmology, with some models reaching a pure de Sitter phase at the present time. Some model parameters can still reproduce well the predictions of $\Lambda$CDM at low redshits.

Significant differences appear in the redshift variation of the ordinary matter energy density, presented in the left panel of Fig.~\ref{fig2}. Interestingly enough, the $\Lambda$CDM model predicts a larger ordinary matter content than the considered model of the two scalar field version of $f(R,T)$ gravity theory, with the differences increasing at higher redshifts. The scalar field $\phi$ is a monotonically decreasing function of the redshift, and its evolution shows a strong dependence on the potential parameters.

\begin{figure}[htbp]
	\centering
	\includegraphics[scale=0.65]{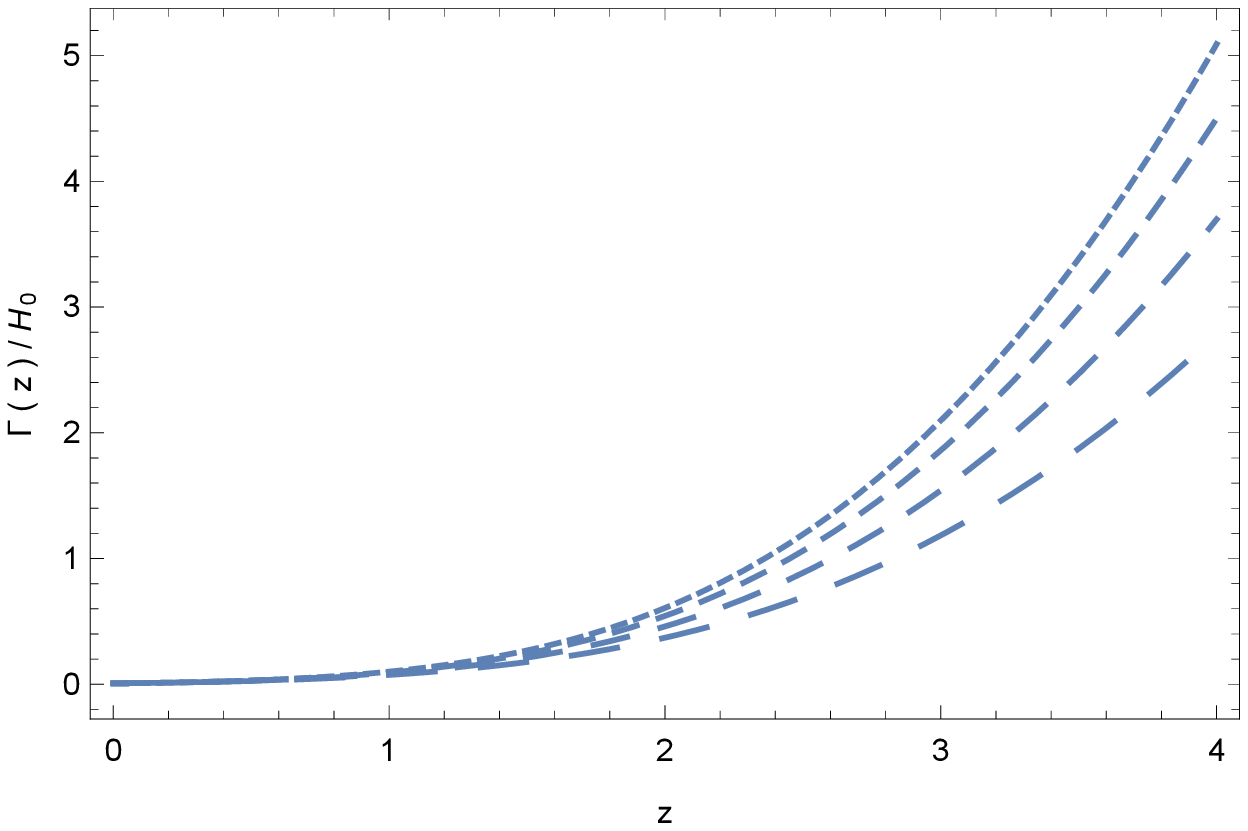}
	\caption{The evolution of the matter creation rate  $\Gamma / H_0$  as a function of redshift $z$ for $\xi=1.54$ (short dashed curve), $\xi=1.61$ (dashed curve), $\xi =1.69$ (long dashed curve), and $\xi=1.76$ (ultra long dashed curve). The values of the potential parameters $\sigma$ and $\epsilon$ have been fixed to $\sigma =-2.89$ and $\epsilon=0.015$, respectively. The initial conditions used for the integration of the cosmological equations are $h(0)=1$, $r(0)=1$ and $\varphi (0)=-0.005$.}\label{fig3}
\end{figure}

The particle creation rate $\Gamma$, normalized to $H_0$, can be obtained as
\be
\frac{\Gamma}{H_0}=\frac{9\epsilon}{2}\frac{hr}{1+(5\epsilon /2)r}.
\ee
For high densities satisfying the condition $(5\epsilon /2)r \gg 1$, $\Gamma /H_0\approx (9/5)h$, and is independent of the matter density, depending only on the rate of the expansion of the Universe. In the opposite limit $(5\epsilon /2)r \ll 1$, $\Gamma /H_0\approx (9\epsilon/2)hr$, and the matter creation rate depends on both the expansion rate and the matter density. The variation of $\Gamma /H_0$ with respect to the redshift is represented in Fig.~\ref{fig3}.

The particle creation rate is a monotonically increasing function of the redshift, indicating a monotonic decrease in time. For redshifts in the range $z\leq 1$,  the creation rate is constant, with matter creation triggering an accelerated expansion of the Universe.

\subsection{$V(\varphi, \psi)=\alpha \varphi ^n \psi^{-m}+\beta \varphi$}

As a second cosmological model in the two scalar field representation of $f(R,T)$ gravity, we will consider the case in which the scalar potential is given by
\be\label{70}
V(\varphi, \psi)=\alpha \varphi ^n \psi^{-m}+\beta \varphi,
\ee
where $\alpha$, $\beta$, $n$ and $m>0$ are constants. In the following we restrict again our analysis to the pressureless case, $p=0$. For these choices of $V$ and $p$ the second equation from Eqs.~(\ref{Fr3}) gives immediately
\be
\rho=\alpha m \frac{\varphi ^n}{\psi ^{m+1}}.
\ee
By introducing the set of dimensionless variables $\left(r,h,\tau, \alpha _0,\beta _0\right)$, defined by $\rho =\left(3H_0^2/8\pi\right)r$, $H=H_0h$, $\tau =H_0t$, $\alpha =\left(3H_0^2/8\pi\right)\alpha _0$, and $\beta =\left(3H_0^2/8\pi\right)\beta _0$, respectively, we obtain
\be\label{71}
r=\alpha _0m\frac{\varphi ^n}{\psi^{m+1}}.
\ee
The first of the Eqs.~(\ref{Fr3}) gives for $h$ the temporal evolution equation
\be
\frac{dh}{d\tau}=\frac{n\alpha _0}{16\pi}\frac{\varphi^{n-1}}{\psi ^m}+\frac{\beta _0}{16\pi}-2h^2.
\ee
The time variation of $\varphi$ can be obtained from Eq.~(\ref{Fr1}) as
\be
h\frac{d\varphi}{d\tau}=\alpha _0m\frac{\varphi ^n}{\psi^{m+1}}+\frac{(1+3m)\alpha _0}{16\pi}\frac{\varphi ^n}{\psi^m}+\frac{\beta _0}{16\pi}-h^2\varphi.
\ee

Equation~(\ref{Fr4}) can be reformulated in a dimensionless form as
\be
\left(1+\frac{3}{16\pi}\psi\right)\frac{dr}{d\tau}+3\left(1+\frac{1}{8\pi}\psi\right)hr+\frac{1}{8\pi}r\frac{d\psi}{d\tau}=0.
\ee

 With the use of the expression (\ref{71}) for $r$ we obtain the evolution equation of $\psi$ as
 \be
 \frac{d\psi}{d\tau}=\frac{\psi \left[6 h (\psi +8 \pi ) \varphi +n (3 \psi
   +16 \pi ) (d\varphi /d\tau)\right]}{\varphi  \left[(3 m+1) \psi +16 \pi  (m+1)\right]}.
 \ee

 Hence, the full system of equations describing the cosmological evolution in the redshift space of the two scalar field $f(R,T)$ gravity theory with potential (\ref{70}) is given by
 \be\label{m4}
-(1+z)h(z) \frac{dh(z)}{dz}=\frac{n\alpha _0}{16\pi}\frac{\varphi^{n-1}(z)}{\psi ^m(z)}+\frac{\beta _0}{16\pi}-2h^2(z),
 \ee
 \bea\label{m5}
-(1+z) h^2(z)\frac{d\varphi (z)}{dz}=\frac{\beta _0}{16\pi}\varphi[z]
-h^2(z)\varphi (z)
	\nonumber \\
+\alpha _0m\frac{\varphi ^n(z)}{\psi^{m+1}(z)}+\frac{(1+3m)\alpha _0}{16\pi}\frac{\varphi ^n(z)}{\psi^m(z)},
 \eea
 and
\begin{eqnarray}\label{m6}
&&-(1+z)h(z) \frac{d\psi (z)}{dz}=\frac{\psi (z)}{\varphi (z)}
\Big[ 6h(z)(\psi (z)+8\pi )\varphi(z)
	\nonumber  \\
&& \qquad \qquad -n(3\psi (z)+16\pi )(1+z)h(z)\frac{d\varphi (z)}{dz}\Bigg]
	\nonumber \\
&& \qquad \qquad \qquad \big/\left[ (3m+1)\psi (z)+16\pi (m+1)\right] ,
\end{eqnarray}
%
respectively. The system of equations (\ref{m4})--(\ref{m6}) must be integrated with the initial conditions $h(0)=1$, $\varphi (0)=\varphi _0$, and $\psi (0)=\psi _0$, respectively. However, the initial values $\phi_0$ and $\psi _0$ are not independent, since the condition $r(0)=1$ gives the relation
\be
\psi (0)=\left(\alpha _0m\varphi _0^n\right)^{1/(m+1)}.
\ee

\begin{figure*}[htbp]
	\centering
	\includegraphics[scale=0.65]{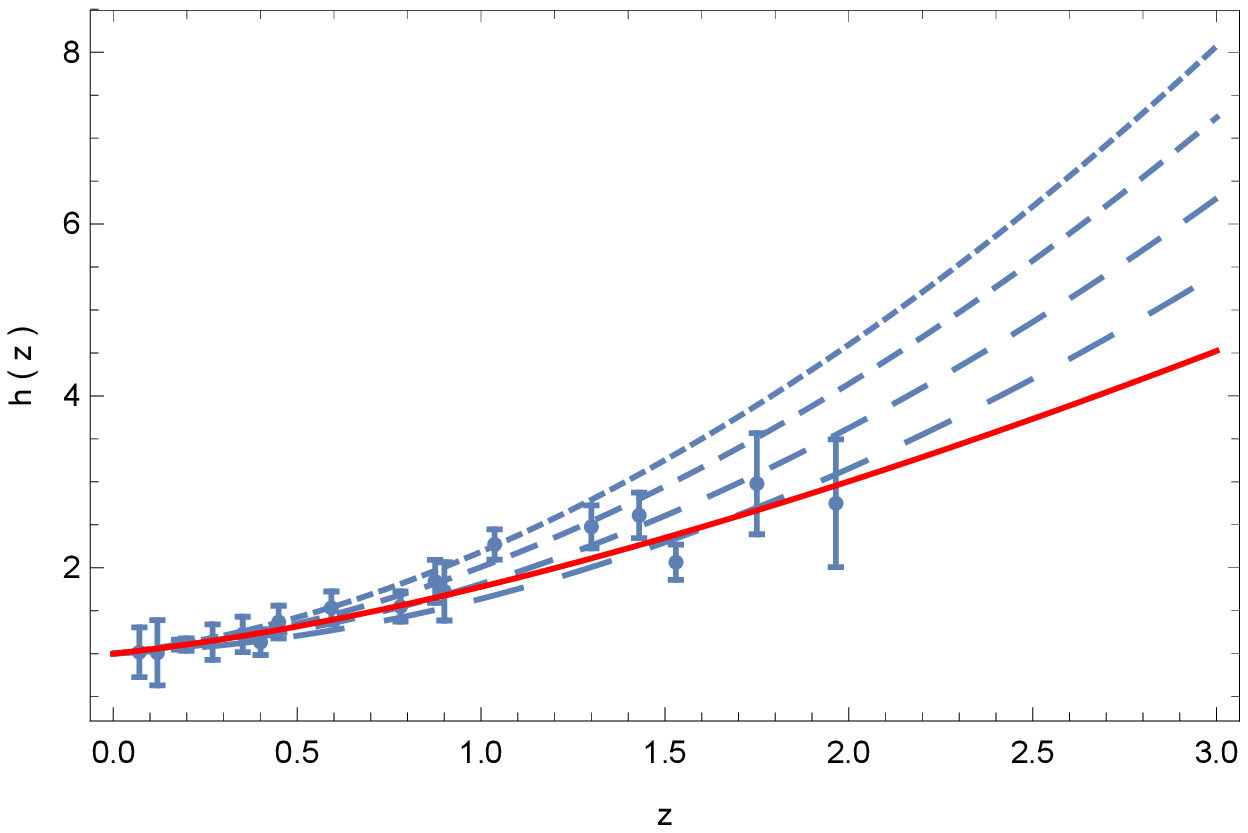}
\includegraphics[scale=0.65]{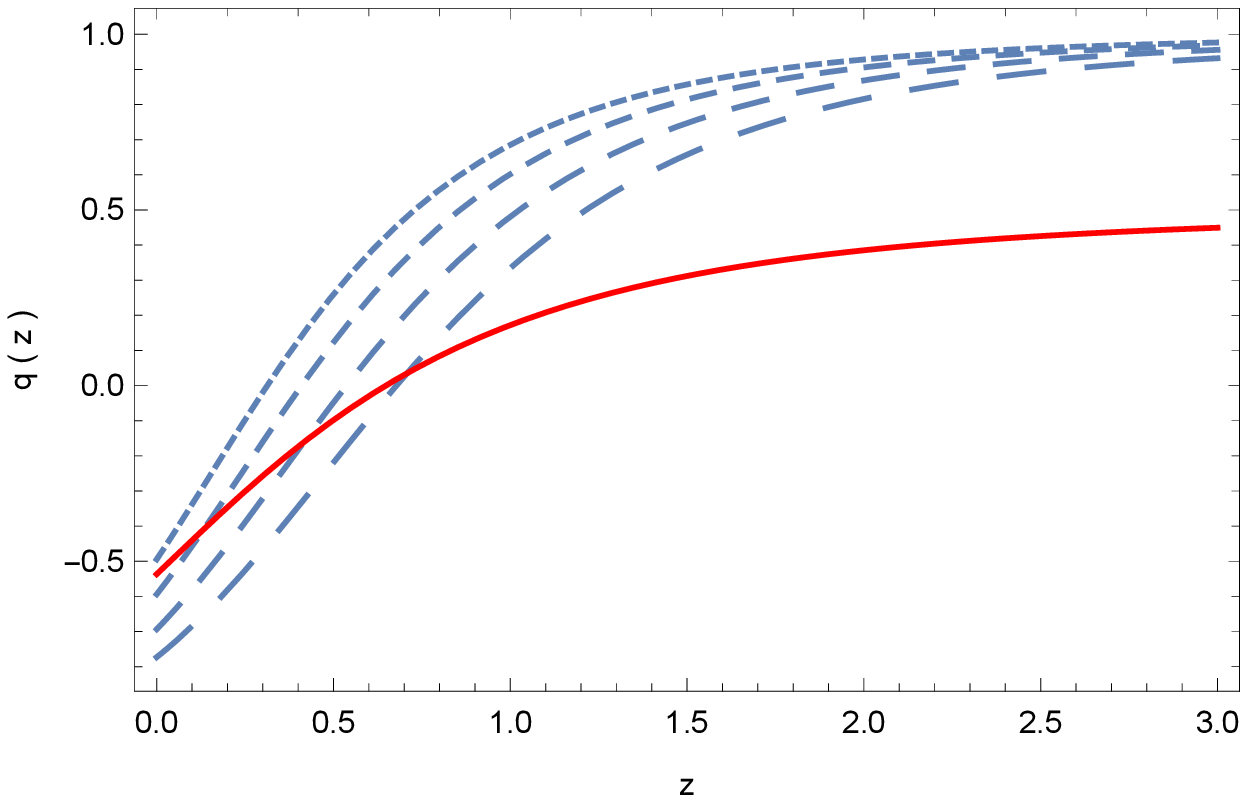}
	\caption{The evolution of the Hubble function $h(z)$  (left panel), and of the deceleration parameter $q(z)$ (right panel)  as a function of the redshift $z$ for the potential $V(\varphi, \psi)=\alpha \varphi ^n \psi^{-m}+\beta \varphi$ for  $\beta_0=75$ (short dashed curve), $\beta_0=80$ (dashed curve), $\beta _0 =85$ (long dashed curve), and $\beta _0=89$ (ultra long dashed curve), respectively. The values of the potential parameters $\alpha _0$, $n$ and $m$ have been fixed to $\alpha _0 =0.1$, $n=1$, and $m=2$, respectively. The initial conditions used for the integration of the cosmological equations are $h(0)=1$, $\varphi (0)=3.90$, and $\psi (0)=\left(\alpha _0 m\varphi _0^n\right)^{1/(m+1)}$.  The solid red curve depicts the predictions of the $\Lambda$CDM model.  The error bars in the left panel indicate the observational values of the Hubble function \cite{H1,H2}.}\label{fig6}
\end{figure*}

The redshift evolution of the Hubble function and of the deceleration parameter are presented, for fixed values of $\alpha _0$, $\phi_0$, $\psi _0$, $m$
and $n$, and for different values of $\beta _0$, in Fig.~\ref{fig6}.
As one can see from the left panel of Fig.~\ref{fig6}, at low redshifts the model can describe well the observational data, and also reproduces almost exactly the predictions of the $\Lambda$CDM model. However, at higher redshifts the present model predicts higher rates for the Hubble function, as compared to the $\Lambda$CDM model. The behavior of the deceleration parameter, represented in the right panel of Fig.~\ref{fig6}, also shows major differences with respect to $\Lambda$CDM at high redshifts, the deceleration rate of the Universe being generally higher in the two scalar field representation of the $f(R,T)$ theory with the potential (\ref{70}). At low redshifts for some particular values of $\beta _0$ one can reproduce the deceleration parameter behavior of $\Lambda$CDM.

\begin{figure*}[htbp]
	\centering
	\includegraphics[scale=0.65]{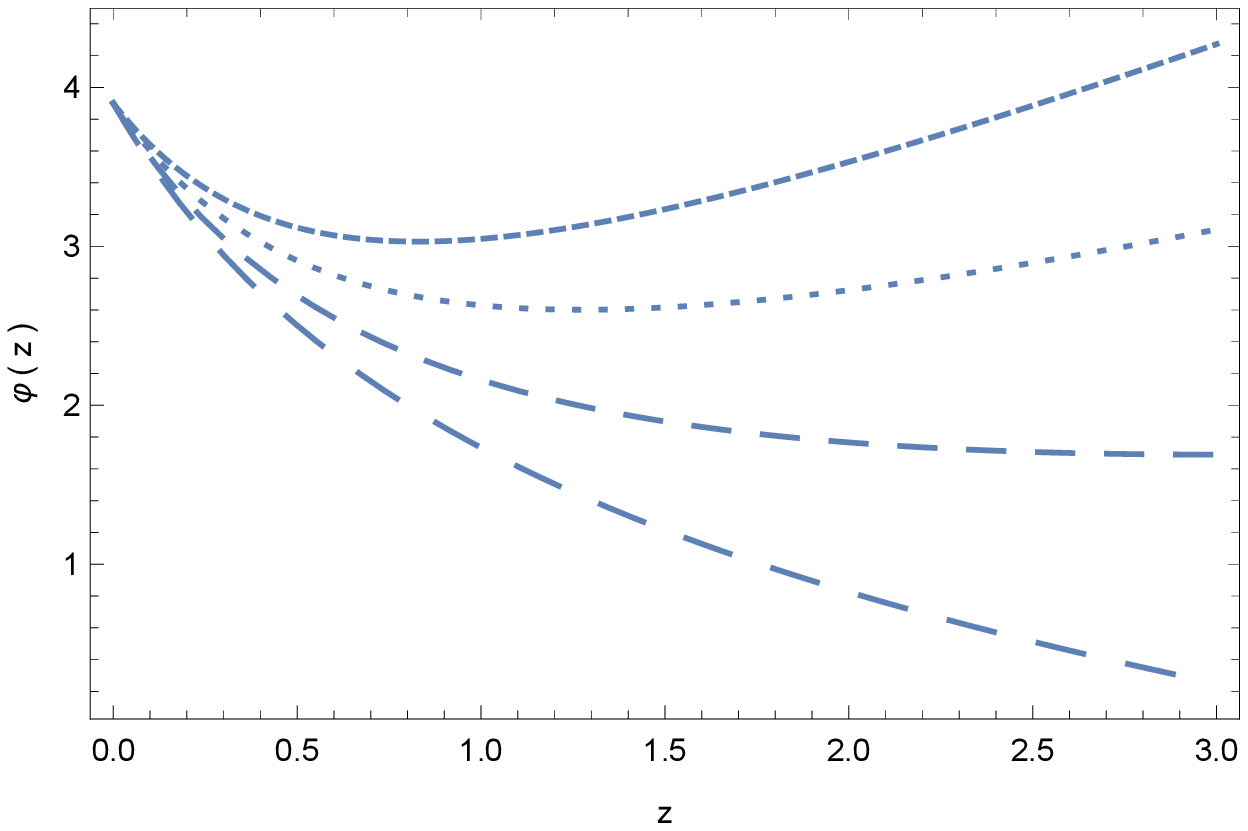}
\includegraphics[scale=0.65]{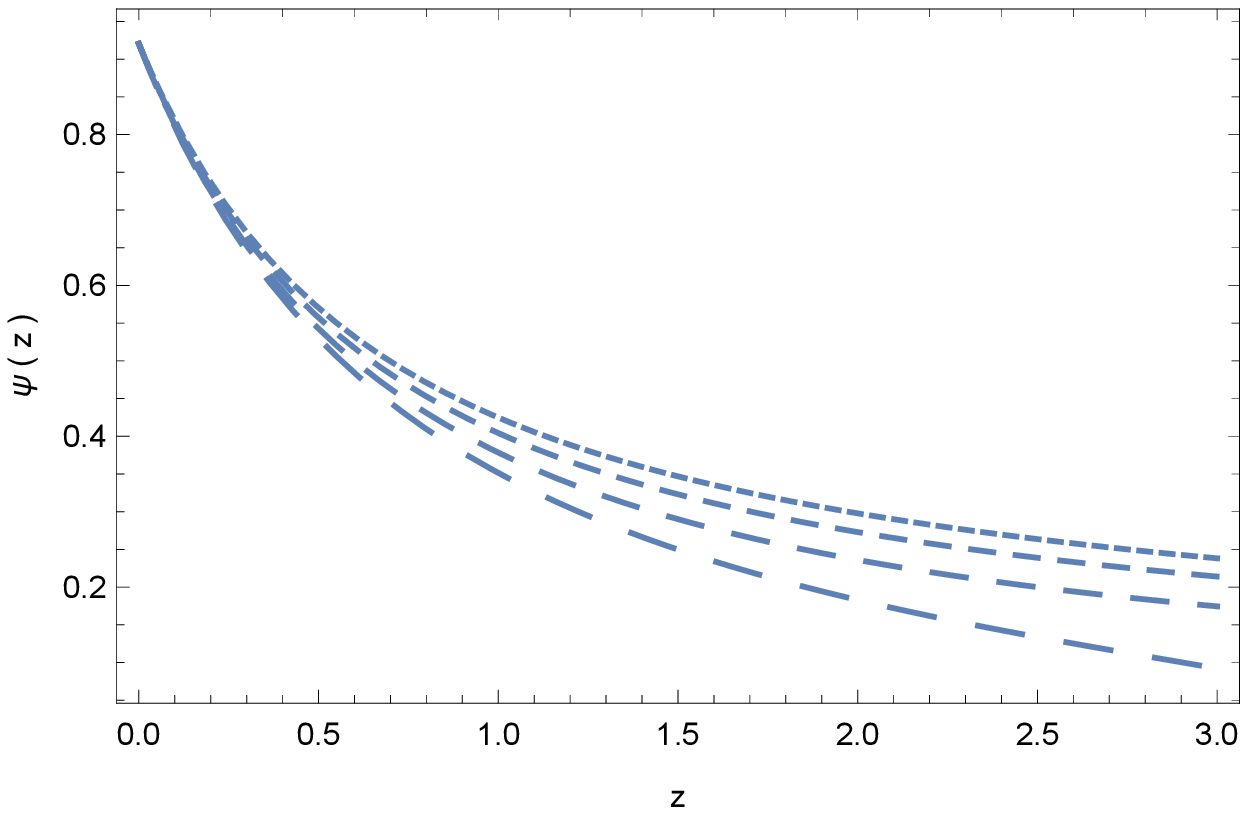}
	\caption{Redshift evolution of the functions $\varphi(z)$  (left panel) and  $\psi (z)$ (right panel) for the potential $V(\varphi, \psi)=\alpha \varphi ^n \psi^{-m}+\beta \varphi$ and for $\beta_0=75$ (short dashed curve), $\beta_0=80$ (dashed curve), $\beta _0 =85$ (long dashed curve), and $\beta _0=89$ (ultra long dashed curve), respectively. The values of the potential parameters $\alpha _0$, $n$ and $m$ have been fixed to $\alpha _0 =0.1$, $n=1$, and $m=2$, respectively. The initial conditions used for the integration of the cosmological equations are $h(0)=1$, $\varphi (0)=3.90$, and $\psi (0)=\left(\alpha _0m\varphi _0^n\right)^{1/(m+1)}$.   }\label{fig7}
\end{figure*}

The variations of the functions $\varphi(z)$ and $\psi (z)$ are represented in Fig.~\ref{fig7}.
The function $\varphi (z)$ has a strong dependence on the model parameters. For smaller values of $\beta _0$, after an initial period of decrease, and after reaching a minimum value, $\varphi (z)$ begins to increase rapidly. For larger values of $\beta _0$, $\varphi (z)$ is a monotonically decreasing function of the redshift (a monotonically increasing function of time). The function $\psi (z)$ decreases monotonically with the redshift, and the variations of $\beta _0$ do not affect the qualitative behavior of $\psi$.

\begin{figure*}[htbp]
	\centering
	\includegraphics[scale=0.65]{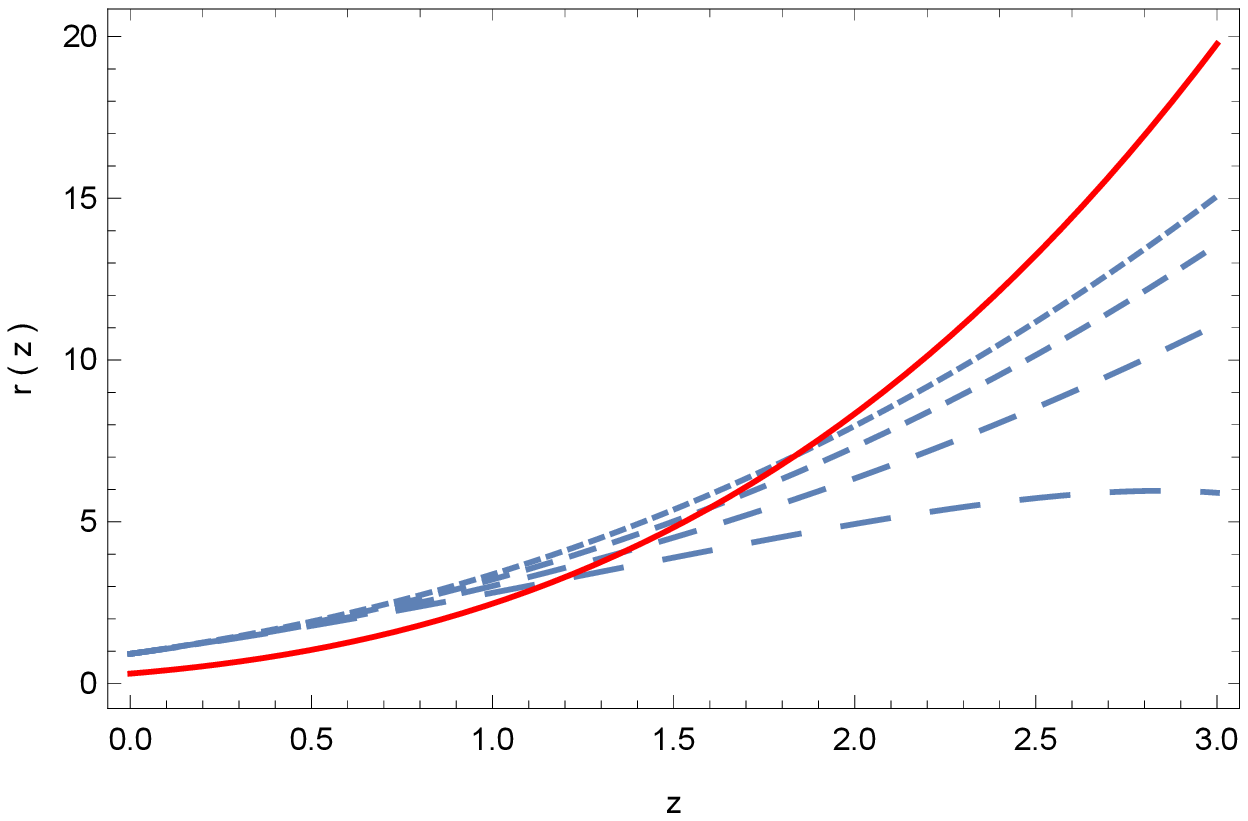}
\includegraphics[scale=0.65]{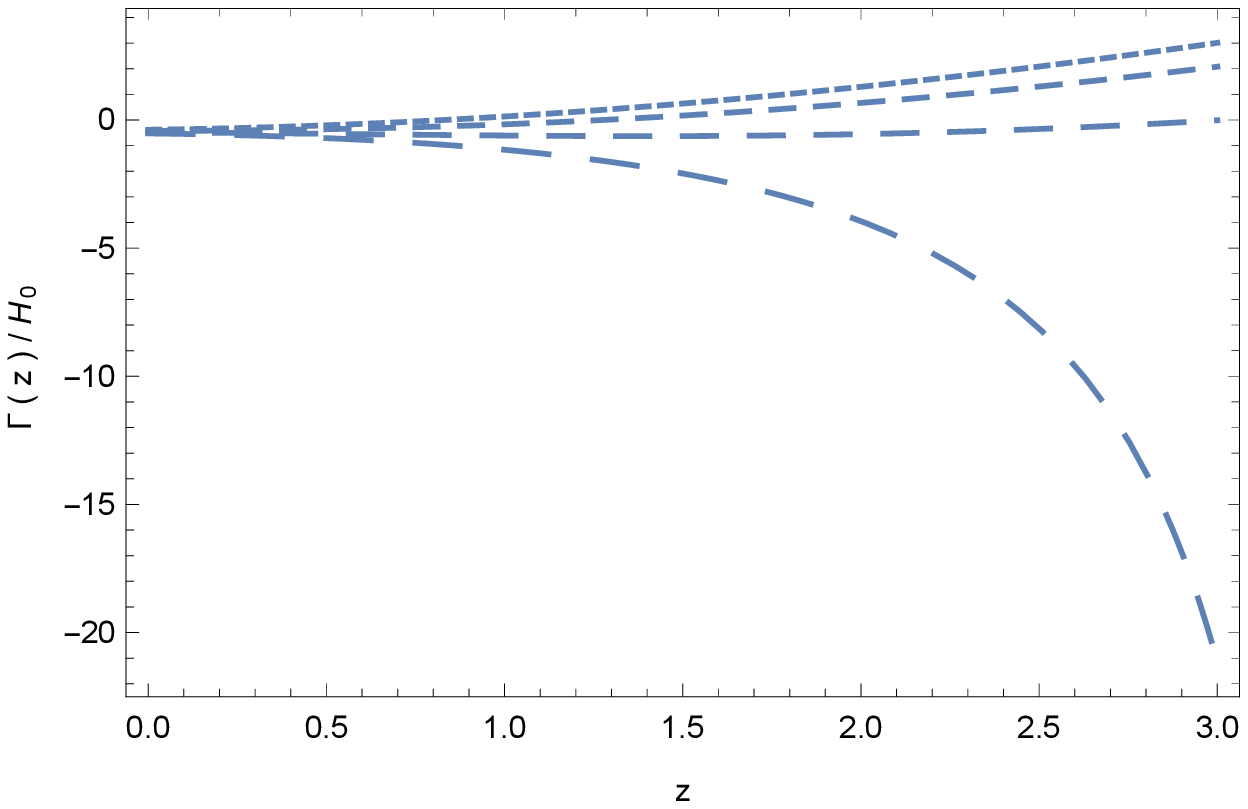}
	\caption{Redshift evolution of the matter density $r(z)$  (left panel) and  of the particle creation rate $\Gamma (z)$ (right panel) for the potential $V(\varphi, \psi)=\alpha \varphi ^n \psi^{-m}+\beta \varphi$ and for $\beta_0=75$ (short dashed curve), $\beta_0=80$ (dashed curve), $\beta _0 =85$ (long dashed curve), and $\beta _0=89$ (ultra long dashed curve), respectively. The values of the potential parameters $\alpha _0$, $n$ and $m$ have been fixed to $\alpha _0 =0.1$, $n=1$, and $m=2$, respectively. The initial conditions used for the integration of the cosmological equations are $h(0)=1$, $\varphi (0)=3.90$, and $\psi (0)=\left(\alpha _0m\varphi _0^n\right)^{1/(m+1)}$. The solid red curve indicates the predictions of the $\Lambda$CDM model.  }\label{fig8}
\end{figure*}

The redshift variations of the matter density of the matter creation rate are shown in Fig.~\ref{fig8}.
The present model gives an acceptable description of the variation of the matter density, and at low redshifts one can reproduce the $\Lambda$CDM model on a qualitative level. At high redshifts the matter density increases more slowly in the present model, and the amount of baryonic matter in the Universe is smaller than the one predicted by standard cosmology.

\section{Discussion and Conclusions}\label{sec:conclusion}

In the present work, we have explored the possibility of gravitationally generated particle production in the scalar-tensor representation of the $f(R,T)$ gravity theory. Due to the explicit nonminimal curvature-matter coupling in the theory, the divergence of the matter energy-momentum tensor does not vanish. We have considered the physical and cosmological implications of this property by using the formalism of irreversible thermodynamics of open systems in the presence of matter creation/annihilation. The particle creation rates, the creation pressure, the temperature evolution, and the expression of the comoving entropy were obtained in a covariant formulation, and discussed in detail. Applied together with the gravitational field equations, the thermodynamics of open systems lead to a generalization of the standard $\Lambda$CDM cosmological paradigm, in which the particle creation rates and pressures are effectively considered as components of the cosmological fluid energy-momentum tensor. We also considered specific models, and we have compared the scalar-tensor $f(R,T)$ cosmology with the $\Lambda$CDM scenario, as well as with the observational data for the Hubble function. The properties of the particle creation rates, of the creation pressures, and the entropy generation through gravitational matter production in both low and high
redshift limits were investigated in detail.
From a cosmological point of view, the generalized Friedmann equations of the scalar-tensor representation of the $f(R,T)$ gravity have the important property of admitting a de Sitter type solution, which leads to the possibility of an immediate explanation of the present day acceleration of the Universe. The de Sitter solution can either describe a constant density Universe, or a Universe in which the matter density decreases asymptotically as an exponential function.

It would also be important to obtain a qualitative estimate of the particle production rate that could explain the accelerated de Sitter expansion of the Universe, without resorting to the presence of dark energy. Such an estimate of the particle production rate can be obtained as follows. As a result of the expansion of the Universe, during the de Sitter phase, the matter density decreases as $\rho (t) = \rho_0 \exp\left(-3H_0t\right)$, where $\rho_0$ denotes the present matter density of the Universe. Consequently, the
variation of the density with respect to the time is given by
\be
\dot{\rho}(t) = -3H_0\rho_0 \exp\left(-3H_0t\right).
\ee
We estimate now this relation at the present time. Moreover, we assume that $\rho_0$ is equal to the critical density of the Universe, $\rho_0 = 3H_0^2 /8\pi G$. Then, for the present day time derivative of the density  we obtain
\be
\left.\dot{\rho} (t)\right|_{t=0} = -\frac{9H_0^3} {8\pi G}.
\ee
Consequently, the particle creation rate necessary to maintain the matter density constant is
\be
\left.\Gamma (t)\right|_{t=0} = -\left.\dot{\rho} (t)\right|_{t=0} = \frac{9H_0^3} {8\pi G}.
\ee

If the creation rate has the above expression, the evolution of the Universe is of de Sitter type, and the ordinary matter density is a constant, satisfying the relation  $3H_0^2 = 8\pi G \rho _{0} = {\rm constant}$. Let’s estimate now the numerical value of matter creation rate $\Gamma$. By assuming  $H_0 = 2.2 \times 10^{-18}\; {\rm s}^{-1}$ (Planck data),
we obtain $\Gamma  = 5.71 \times 10^{-47} \;{\rm g/cm}^3/{\rm s}$. We convert now cm to km and seconds to years, respectively, thus obtaining $\Gamma  = 1.8 \times
10^{-24}\;{\rm  g/km}^3/{\rm year}$. This result shows that the creation of a single proton in one km$^3$ in one year, or, equivalently,  160 protons in a km$^3$ in a century, can fully balance the decrease in the density of the matter due to the de Sitter evolution. Of course, such a small amount of matter, created presumably from vacuum due to quantum field theoretical processes, cannot be detected observationally or experimentally.

In concluding, the formalism of irreversible matter creation of thermodynamics of open systems as applied in cosmology can give a full account of the creation of matter in a homogeneous and isotropic Universe, as long as the particle creation rate, and consequently the creation pressure, are not zero. For instance, Einstein's GR is incapable of explaining the increase in entropy that accompanies matter creation, since both the creation rate, and the creation pressure, are not of gravitational origin. Modified theories of gravity in which these two quantities do not vanish not only can provide a (macroscopic) phenomenological description of particle production in the cosmological fluid filling the Universe but also lead to the possibility of cosmological models that start from empty conditions and gradually build up matter and entropy. Hence,  $f(R,T)$ gravity theory can provide a phenomenological description of the matter creation processes in the Universe. In this theory the geometry-matter coupling is responsible for inducing particle production from the gravitational field.

\begin{acknowledgments}
The work of TH is supported by a grant of the Romanian Ministry of Education and Research, CNCS-UEFISCDI, project number PN-III-P4-ID-PCE-2020-2255 (PNCDI III). FSNL acknowledges support from the Funda\c{c}\~{a}o para a Ci\^{e}ncia e a Tecnologia (FCT) Scientific Employment Stimulus contract with reference CEECINST/00032/2018, and funding through the research grants UIDB/04434/2020, UIDP/04434/2020, PTDC/FIS-OUT/29048/2017, CERN/FIS-PAR/0037/2019 and PTDC/FIS-AST/0054/2021.
\end{acknowledgments}



\end{document}